\documentclass[usenatbib]{mn2e}
\usepackage{times}
\usepackage{graphicx} 
\usepackage{amsmath}
\usepackage{mathabx}
\usepackage{natbib}

\begin{document}
\title[The GOGREEN Survey]{Gemini Observations of Galaxies in Rich Early Environments (GOGREEN) I: Survey Description}
\author[Balogh et al.]{Michael L. Balogh$^{1}$\thanks{mbalogh\@@uwaterloo.ca}, David G. Gilbank$^{2,3}$, Adam Muzzin$^{4}$, Gregory Rudnick$^{5}$, 
\newauthor	 Michael C. Cooper$^{6}$, Chris Lidman$^{7}$, Andrea Biviano$^{8}$, Ricardo Demarco$^{9}$, Sean L. McGee$^{10}$, 
\newauthor Julie B. Nantais$^{11}$, Allison Noble$^{12}$, Lyndsay Old$^{13}$, Gillian Wilson$^{14}$, Howard K.C. Yee$^{13}$,  
\newauthor  Callum Bellhouse$^{10,15}$, Pierluigi Cerulo$^{9}$, Jeffrey Chan$^{14}$, Irene Pintos-Castro$^{13}$, 
\newauthor Rane Simpson$^{1}$, Remco F. J. van der Burg$^{16}$, Dennis Zaritsky$^{17}$, Felicia Ziparo$^{10}$, 
\newauthor M. Victoria Alonso$^{18}$, Richard G. Bower$^{19,20}$, Gabriella De Lucia$^{8}$, Alexis Finoguenov$^{21,22}$, 
\newauthor Diego Garcia Lambas$^{18}$, Hernan Muriel$^{18}$,  Laura C. Parker$^{23}$, Alessandro Rettura$^{24}$, 
\newauthor Carlos Valotto$^{18}$, Andrew Wetzel$^{25,26,27}$\thanks{Caltech-Carnegie Fellow}
\\
\\Author affiliations are listed at the end of the paper
}
\date{\today}
\maketitle
\begin{abstract}
We describe a new Large Program in progress on the Gemini North and South telescopes: Gemini Observations of Galaxies in Rich Early Environments (GOGREEN). This is an imaging and deep spectroscopic survey of 21 galaxy systems at $1<z<1.5$, selected to span a factor $>10$ in halo mass.  The scientific objectives include measuring the role of environment in the evolution of low-mass galaxies, and measuring the dynamics and stellar contents of their host haloes. The targets are selected from the SpARCS, SPT, COSMOS and SXDS surveys, to be the evolutionary counterparts of today's clusters and groups.  The new red-sensitive Hamamatsu detectors on GMOS, coupled with the nod-and-shuffle sky subtraction, allow simultaneous wavelength coverage over $\lambda\sim 0.6$--$1.05\mu$m, and this enables a homogeneous and statistically complete redshift survey of galaxies of all types.  The spectroscopic sample targets galaxies with AB magnitudes $z^{\prime}<24.25$ and [3.6]$\mu$m$<22.5$, and is therefore statistically complete for stellar masses  $M_\ast\gtrsim10^{10.3}M_\odot$, for all galaxy types and over the entire redshift range.  Deep, multiwavelength imaging has been acquired over larger fields for most systems, spanning $u$ through $K$, in addition to deep IRAC imaging at 3.6$\mu$m.  The spectroscopy is $\sim 50$ per cent complete as of semester 17A, and we anticipate a final sample of $\sim 500$ new cluster members.  Combined with existing spectroscopy on the brighter galaxies from GCLASS, SPT and other sources, GOGREEN will be a large legacy cluster and field galaxy sample at this redshift that spectroscopically covers a wide range in stellar mass, halo mass, and clustercentric radius.  
\end{abstract} 
\section{Introduction}

Galaxy clusters are extraordinarily valuable as laboratories for a wide range of tests and experiments.  They play a central role in studies of cosmology \citep[e.g.][]{CNOC1,WS98,HMH,ACT_cc,SPT_SZcc}, galaxy and structure formation \citep[e.g.][]{Dressler,BO84,PSG,Erica,vdB07,PLRC, Wetzel+13}, high energy physics \citep[e.g.][]{MME,TN01,PE04,BEM,entropy,CDVS,Perseus00,Hitomi} supermassive black hole growth \citep[e.g.][]{MN_review,AH12}, and for the determination of the nature of dark matter \citep[e.g.][]{bullet,sidm_bullet,Bradac+08,J+12}.  Their enormous gravitational potentials allow them to act as cosmic ``calorimeters'', maintaining an observable record of all the energy inputs and outputs associated with galaxy formation over the history of the Universe \citep[e.g.][]{VB01,GZZ,BMBE,Vik+05,PS08}.  They host the most massive galaxies, whose stars are among the first to form \citep[e.g.][]{Lidman_bcg,Lin+13,MM+14,Fassbender14}.  Clusters are also the ideal places to study rare and extraordinary perturbations to galaxy evolution, such as hydrodynamic stripping of gas \citep[e.g.][]{Vollmer00,Merluzzi+12,Boselli+16}, tidal stripping of matter \citep{Nat+02}, and high-speed gravitational encounters.  Much of what we have learned about galaxy evolution is thanks to years of research on these systems.  

At $z>1$, when the gas accretion rates, relative gas masses and star formation rates of galaxies were much higher than they are today, the interactions between galaxies and their environments are also expected to be very different.  Large spectroscopic samples have now been built up in clusters approaching $z\sim 1$ \citep{GCLASS12,GEEC2-survey,SPT-GMOS} but little is known about the physical properties of typical galaxies in $z>1$ clusters.  Those spectroscopic studies that do exist are generally restricted to the most massive galaxies \citep[e.g.][]{ISCS_rs,Lotz_z162} or emission line galaxies \citep[e.g.][]{ISCS_Ha,ZFIRE} in the most massive clusters \citep[e.g.][]{Nantais+13,Madcows_Martini,Madcows}.  Studying the faint, and more common, galaxy population at $1<z<1.5$ usually relies heavily on photometric redshifts \citep[e.g.][]{SpARCS_Nantais}.

Thus for the first round of Gemini Large and Long Programs (LLP) in 2014 we proposed an ambitious distant cluster legacy survey, titled {\it Gemini Observations of Galaxies in Rich Early ENvironments (GOGREEN)}, using the GMOS instruments on the North and South telescopes.   GMOS has several capabilities that make it ideally suited to studies of galaxy clusters at $z\sim 1$.  The nod-and-shuffle (n\&s) mode \citep{GBH} allows excellent sky subtraction at red wavelengths, resulting in much greater efficiency for faint galaxies \citep[as exploited by the  GDDS\footnote{Gemini Deep Deep Survey}][]{GDDS}.  Moreover, the n\&s microslits are up to three times smaller than normal slits, allowing them to be placed with a very high surface density.  Together with the new Hamamatsu detectors, which have good sensitivity up to $\lambda\sim1.05\mu$m, GMOS spectroscopy of faint objects is now feasible at $1.0<z<1.5$.   

 The objective of the survey is to build directly on the work we have done with Gemini in constructing the GCLASS\footnote{Gemini CLuster Astrophysics Spectroscopic Survey} cluster \citep{GCLASS12} and GEEC2\footnote{Galaxy Environment Evolution Collaboration 2} group \citep{GEEC2-survey} surveys.  GCLASS was a GMOS survey consisting of $\sim 450$ spectroscopically-confirmed members of ten massive clusters at $0.86<z<1.34$.  Among other things, this enabled new insight into the environmental transformation of galaxies \citep{Muzzin14,Foltz15,Noble+13,Noble+16}, the stellar mass content and distribution in clusters \citep{vdB+13,GCLASS_vdB}, cluster dynamics \citep{GCLASS_dynamics} and growth of the brightest cluster galaxies \citep{Lidman_bcg}.  An independent but highly complementary survey was GEEC2, which used a sample of ten X-ray detected groups in the COSMOS field to address similar questions of galaxy evolution \citep{GEEC2-1,Mok13,Mok14} and group dynamics \citep{GEEC2-Hou}.  The combination of GEEC2 and GCLASS provides a sample that spans more than two orders of magnitude in halo mass, allowing the measurement of halo mass effects on the environmental quenching measurement \citep{Balogh+16}.
 
 GOGREEN uses a similar strategy to extend these works to $1.0<z<1.5$ with comparably dense spectroscopy on 21 systems spanning a wide range in halo mass.   An important feature of GOGREEN is that our systems are chosen to be representative of the progenitors of today's clusters; this is complementary to efforts focussed on the most massive clusters at high redshift, which have few, if any, local descendants.  The survey will obtain spectroscopy on a large sample of very faint targets, $z^{\prime}<24.25$ and $[3.6]<22.5$ to obtain a sample of confirmed cluster members, measure cluster dynamics and galaxy stellar populations, and provide critical calibration of photometric redshifts.
The survey design is driven by three key science goals, and an aim to provide a legacy dataset that is useful to the broader community.  These goals are described in more detail below.

\subsection{Environmental quenching and growth of the stellar mass function}
Despite a solid theoretical foundation for the gravitational growth of
dark matter structure, galaxy formation models have great difficulty simultaneously reproducing the rate of decline in global SFR, the mass
dependence of this decline, and the star formation histories of
satellite galaxies \citep{BBC12,Weinmann12,Hirschmann+14,DeL+12,Henriques,Illustris_Genel,EAGLE_Trayford15}.  These problems may be related, as they are all sensitive to assumptions about how gas accretion,  ejection and heating processes depend on epoch, environment and halo mass \citep{McGee14}. 
The conventional picture of the interaction between galaxies and their surroundings is that galaxies enter dense environments with a reservoir of gas (either in the stellar disk, or the halo), and that star formation declines as this reservoir is removed \citep[e.g.][]{infall,Bower05,Schawinski+14,Fillingham15}.  However, cosmological simulations show that galaxies grow as a result of continuous infall from surrounding filaments \citep[e.g.][]{KKWD}, a scenario that is supported by indirect observational arguments \citep[e.g.][]{Dave12,Lilly+13}. While a reservoir may play a role at low redshift, at higher redshift the supply of fresh gas fully dominates over the consumption of the reservoir.  This change leads to a prediction that dense environments shut down star formation even more rapidly at $z>1$ than at low redshift \citep{McGee14,Balogh+16,vdV16}.  The sensitivity of the observed galaxy population to gas accretion and outflow rates on large scales allows us to use trends with environment to put constraints on feedback and accretion models that may be relevant to the evolution of all galaxies.

Simple but powerful indicators of SFR suppression (or ``quenching'') are the evolution of the quiescent galaxy stellar mass function, and the stellar--mass dependence of the quiescent fraction \citep[e.g.][]{vdB07,zCOSMOS_Peng,Fillingham16,Balogh+16}.
From these measurements alone it is possible to put strong constraints on the quenching timescale and its evolution \citep[e.g.][]{TW10}, which is a powerful indicator of how gas-supply and removal mechanisms change with time \citep{McGee14,Balogh+16,Fossati17}.
In the field population the quiescent galaxy mass function evolves rapidly, as star formation is shut down first in the most massive galaxies, and later in dwarfs \citep[e.g.][]{Muzzin_SMF}.  In massive clusters this is clearly seen as the growth of the ``red sequence'' \citep[e.g.][]{EDisCS,Tanaka05,GB08,Rudnick+09,Rudnick+12}.  At $z<1$, most models predict many more low-mass, quiescent galaxies than are observed, a consequence of the well-established overquenching problem \citep{Weinmann+11,Hirschmann+14,DeL+12,Henriques,Illustris_Genel,EAGLE_Trayford15}.  

The situation is much less clear at higher redshift \citep[e.g.][]{SpARCS_Nantais}, where the gas content, accretion rates and star formation rates of galaxies are so much higher, and even galaxies in cluster cores have only been satellites for a few Gyr.    
Moreover, the higher average star formation rate of field galaxies, and the increased rate at which they are accreted by the cluster, translates directly into a much higher fraction of galaxies observed in the ``transition phase'' between actively star--forming and quiescent \citep{Ediscs_psb2,Mok14}.  Exceptional sensitivity to the galaxy transformation timescale can be obtained from fairly straightforward modeling of the radial gradients and projected phase space distribution of such subpopulations, compared with the quiescent and star--forming galaxies \citep[e.g.][]{infall,Erica,McGee-accretion,Noble+13,Taranu,Muzzin14,Haines15}.

GOGREEN is designed specifically to measure the quiescent fraction of galaxies at $1.0<z<1.5$, over a factor $>10$ in halo mass, with a spectroscopic sample statistically complete for all galaxy types down to stellar masses of $M_\ast\sim10^{10.3}M_\odot$.  In addition to the targeted clusters and groups, the survey will result in a comparably--sized field sample selected in the same way.  The deep, multiwavelength imaging ensures a robust and homogeneous separation of passive from star-forming galaxies, and a photometric redshift catalogue that is essential to account for the spatial incompleteness of the spectroscopic sample.  When complete, the total spectroscopic sample size, including bright galaxy spectroscopy from GCLASS and other published catalogues, will be comprised of $\sim 1000$ cluster members; about half of these will be newly acquired via GOGREEN.  This will ensure that statistical uncertainties on the quenched fraction, in bins of stellar and halo mass, are small enough to distinguish between different physical models as described in \citet{Balogh+16}. 

\subsection{The hierarchical assembly of baryons}
It is a fundamental prediction of $\Lambda$CDM theory that massive clusters are built from haloes of lower mass: groups and isolated galaxies \citep[e.g.][]{Berr+08,DeL+12,Bullock-hgf}.  Since it is difficult to preferentially remove stars from dark matter dominated systems, when these systems merge the fraction of total mass in stars can only increase (via star formation) or remain constant.  Therefore, measurement of the stellar fraction, gas fraction, and star formation rate in haloes of a given mass provide one of the closest possible links between galaxies and this basic prediction of the $\Lambda$CDM theoretical framework \citep{Kravtsov2005,GZZ,BMBE,G+09,Gonzalez+13,Leauthaud+11,Leau12}.  Precision measurements of this type are essential for calibrating and constraining models, and are an essential complement to abundance-matching or halo occupation distribution model approaches \citep[e.g.][]{Behroozi+13}. 

With GOGREEN we will directly measure the central and total stellar mass of haloes at  $1.0<z<1.5$.
The spatial and dynamical distribution of cluster galaxies is sensitive not only to the field accretion rate, but also to the dynamical friction time and galaxy merger and disruption timescales.  These rates are not well understood theoretically \citep[e.g.][]{DeLucia_merging}, despite being primarily gravitational processes, and observations of these distributions provide valuable constraints, as we have shown with GCLASS \citep[e.g.][]{GCLASS_vdB}.

\subsection{Cluster Dynamics and Halo Masses}
At low redshift, the total mass content and distribution of galaxy clusters
can be estimated by gravitational lensing, from the properties of the intracluster plasma under the assumption of
hydrostatic equilibrium, or from the
distribution and kinematics of cluster galaxies. The latter method
always provides critical independent information from the other two, and is especially important for clusters at high redshifts, which are
notoriously difficult to detect by their X-ray emission or weak
lensing signal. At $z>1.0$
our knowledge of the mass profiles of galaxy clusters is therefore
limited to only a few individual clusters. 

Dynamical analyses of nearby clusters have shown their $M(r)$ to be well
characterized by either an NFW \citep{NFW}  or an \citet{Einasto} profile, passive galaxy
orbits to be isotropic and star--forming galaxy orbits to be radially elongated \citep[eg.][]{BG03,BK04}. The NFW and
Einasto models appear to also fit well the $M(r)$ of $z\sim 0.6$ clusters,
but the orbits of passive
galaxies evolve with $z$, and at $z\sim 0.6$ are more similar to those
of star--forming galaxies \citep{BP09,Biviano+13}.

Using a stack of $\sim 400$ galaxies in 10 clusters of the GCLASS
sample, \citet{GCLASS_dynamics} have shown that the $M(r)$ of $z \sim 1$
clusters is still well described by the NFW model, with a
concentration as predicted by numerical simulations, and that the
orbits of passive and star-forming cluster galaxies are
indistinguishable and mildly radially elongated.
With GOGREEN we will
trace this evolution to $z\sim 1.5$.
We will be able to measure whether the NFW and Einasto models
remain valid representations of the cluster $M(r)$, which
is particularly interesting as the onset of dynamical equilibrium in
galaxy clusters is still a poorly understood process \citep[e.g.][]{DM05}.   Combined with the velocity anisotropy profile $\beta(r)$ we can measure the more fundamental pseudo-phase space profile \citep{DM05,LC09}.  Evolution in this profile can distinguish between cluster assembly via fast, violent relaxation processes and smooth accretion of matter from the field \citep{Hansen+09}.

\subsection{Legacy Science}
GOGREEN will provide deep, multiwavelength imaging and spectroscopy over 21 systems spanning a factor $>10$ in halo mass.  Future surveys like  eRosita, Euclid and LSST  will find large samples of high-redshift clusters.  These surveys rely on spectroscopic studies to calibrate their observable quantitites in way that is necessary for  cosmological applications.  The depth and completeness of GOGREEN spectroscopy is a good complement to efforts like \citet{SPT-GMOS} and \citet{Madcows} which aim to sparsely sample relatively massive galaxies in a much larger set of clusters.

A byproduct of our survey will be a deep spectroscopic field survey of $>600$ galaxies at $1.0<z<1.5$, with homogeneous and well-understood selection criteria.  At present, none of the existing wide-field spectroscopic surveys have the red sensitivity to match GOGREEN depth at $1.3<z<1.5$.
Claims about evolution in the stellar mass function and star formation history, for example, are based on photometric redshifts, which are notoriously unreliable in regions of parameter space where spectroscopic calibration is unavailable.   The GOGREEN field survey will be twice the size of GDDS \citep{GDDS}, and 0.5 mag deeper, allowing 
an unparalleled spectroscopic measurement of the galaxy mass function, separated by galaxy type.  It will provide a crucial calibration sample for photometric redshifts out to $z=1.5$, needed by surveys like LSST and PanStarrs.

In this paper we describe the survey design (\S~\ref{sec-survey}), spectroscopic observations(\S~\ref{sec-obs}), and the current status of the project (\S~\ref{sec-status}).  All magnitudes reported in this paper are on the AB system. 

\section{Survey Design}\label{sec-survey}
\subsection{Objectives}
GOGREEN is designed primarily to learn about the stellar populations in galaxies that inhabit massive haloes, $M\gtrsim 5\times 10^{13}M_\odot$, at $1.0<z<1.5$.  To do this it is essential to cover a large range in both stellar mass and halo mass.  In particular, it is important to study low stellar-mass galaxies, which are rarely quenched in the field population.  At $1.0<z<1.5$, such galaxies are faint and red, making it challenging to even obtain a redshift since most strong absorption features are redshifted to wavelengths where night sky emission lines are strong.  To take advantage of the range in halo mass, it is important to be able to characterize those haloes, which, in part, requires redshifts for as many cluster members as possible, including the brightest ones.  These two goals --- very deep spectroscopy of faint galaxies, together with a large number of redshifts for bright galaxies --- are difficult to achieve, and most cluster surveys aim to do one or the other.  GOGREEN is specifically designed to achieve both goals within the same program. 
\begin{table}
	\begin{tabular}{lllll}
		Name & RA & Dec & z &$(z^{\prime}-[3.6])_{RS}$\\
		&\multicolumn{2}{c}{(J2000)}&&(AB)\\
		\hline
		\multicolumn{5}{c}{Massive SPT Clusters}\\
		SPT-CL J0205-5829 &31.43900	&-58.48290	&1.320&2.61\\
		SPT-CL J0546-5345 &86.65616	&-53.75800	&1.067&2.01\\
		SPT-CL J2106-5844 &316.51912 &-58.74110	&1.132&2.17\\ 
		\hline
		\multicolumn{5}{c}{SpARCS Clusters}\\
		SpARCS0035-4312	  &8.95708	&-43.20678	&1.335&3.0\\
		SpARCS0219-0531	  &34.93156 &-5.52494	&(1.3)&2.5\\
		SpARCS0335-2929   &53.76487	&-29.48219	&1.368&3.1\\
		SpARCS1033+5753	  &158.35650 &57.89000	&1.455&3.0\\
		SpARCS1034+5818	  &158.70599 &58.30917	&(1.4)&3.0\\
		SpARCS1051+5818	  &162.79680 &58.30087	&1.035&1.9\\
		SpARCS1616+5545	  &244.17180 &55.75714	&1.156&2.2\\ 
		SpARCS1634+4021	  &248.64751 &40.36433	&1.177&2.3\\
		SpARCS1638+4038	  &249.71517 &40.64525	&1.196&2.5\\
		\hline
		\multicolumn{5}{c}{Groups}\\
		SXDF60XGG		  &34.18937	&-5.16353	&1.410&3.29\\
		SXDF64XGG	      &34.32375	&-5.17140	&1.030&2.41\\
		SXDF87XGG	      &34.52729	&-5.05699	&1.402&2.81\\ 
		SXDF49XGG	      &34.53474	&-5.07140	&1.059&2.20\\
		SXDF76XGG	      &34.74128	&-5.32334	&(1.4)&3.01\\
		COSMOS-28	      &149.45758 &1.67241	&1.258&2.5\\ 
		COSMOS-63	      &150.35332 &1.93337	&1.234&2.45\\
		COSMOS-221	      &150.57024 &2.49864	&1.146&2.41\\
		COSMOS-125	      &150.62720 &2.15920	&(1.45)&3.21\\
		\hline
	\end{tabular}
	\caption{The table presents the 21 galaxy clusters and groups in the GOGREEN sample, in ordered by RA within three mass classes.  Redshifts are given in column (4); values in parentheses are estimates.  The final column gives the ($z^{\prime}$-[3.6]) colour of the identified red sequence, used in mask design (see \S~\ref{sec-masks}).  \label{tab-cluster_sample}}
	
\end{table}

\subsection{Cluster sample}
GOGREEN is constructed to enable robust measurements of the populations and dynamics of cluster members at $1.0<z<1.5$, as a function of  cluster-centric radius and stellar mass.
The greatest power of the survey will come from combining the sample with comparable data on lower redshift systems, such as EDisCS \citep{Ediscs-survey}, MeNEACS \citep{Meneacs}, CCCP \citep{CCCP}, CNOC \citep{YEC}, GEEC \citep{CNOC2_groupsI,GEEC_sedfit} and CLASH \citep{CLASH,CLASH_VLT}.  
These surveys cover a halo mass range $\sim 10^{13}M_\odot$--$\sim5\times10^{15}M_\odot$, for $z<1$.  
In order to sample the antecedents of the lower redshift systems, we select galaxy systems in three approximate bins of richness: groups ($M<10^{14}M_\odot$), typical clusters ($10^{14}<M/M_\odot<5\times 10^{14}$), and very massive clusters ($M>5\times 10^{14}M_\odot$).  The initial focus of our spectroscopy is on the typical and massive clusters, with the groups at a lower priority until we are assured the total time available is not unduly compromised by weather loss. 

\begin{table*}
	\begin{tabular}{lllllll}
		Target &Date&Telescope/ &Integration &PA &Depth&Conditions/\\
		&       &Detector   & time (ks)  & (deg)   &$5\sigma$ (AB)&Notes           \\
		\hline
		SPT0205 & Sept 28, 2014 &GS/Ham & 5.4 &90&25.2& IQ 0.7\arcsec, (1) \\
		SPT0546 & Oct 1, 13, 14, 2014 & GS/Ham & 7.4 &0&24.8& IQ 0.7\arcsec, (1,2) \\
		SPT2106 & April 11, 2015 & GS/Ham & 5.4 &100&24.1&IQ 0.6\arcsec,(1)\\
		SpARCS0035 & Sept 28, 2014 & GS/Ham & 5.4 &0&24.75& IQ 0.65\arcsec, (1) \\
		SpARCS0219 & Oct 14, 2014 & GS/Ham & 5.4 &0&25.0& IQ 0.65\arcsec, (1) \\
		SpARCS0335 & Sept 28, 2014 & GS/Ham & 5.4 &185&25.6& IQ 0.65\arcsec, (1) \\
		SpARCS1033 & March 29, 2015 & GN/EEV & 8.91 &0&25.9&IQ 0.75\arcsec \\
		SpARCS1034 & March 28, 2015 & GN/EEV & 8.91 &90&25.4&IQ 0.7\arcsec \\
		SpARCS1051 & May 8 \& 14, 2015 & GN/EEV & 8.91 &90&25.1& IQ 0.8\arcsec\\
		SpARCS1616 & May 14\&15, 2015 & GN/EEV & 8.91 &0&25.6& IQ 0.8\arcsec\\
		SpARCS1634 & May 14, 2015 & GN/EEV & 5.22 &215&25.7& IQ 0.7\arcsec\\
		SpARCS1638 & May 14\&15, 2015 & GN/EEV & 8.91 &90&25.6& IQ 0.7\arcsec\\
	\end{tabular}
	\caption{The GMOS $z^{\prime}$-band images acquired as part of GOGREEN are described in this table.  Images taken on GMOS-N, with the older EEV detector (GN/EEV in column 3), required longer integration times to accommodate the lower sensitivity, compared with the Hamamatsu detectors on GMOS-S (GS/Ham).  Depths in column (6) are based on analysis of the Gemini standard preimaging pipeline reduction.  Position angles (column 5) are chosen to ensure appropriate guide star availability for the MOS follow up. Notes:  (1) Saturated pixels alter the background level across the amplifier. (2) Image quality affected by poor active optics correction. \label{tab-preim}}
\end{table*}

For the cluster sample, it is natural and efficient to build on the existing investment in GCLASS \citep{GCLASS12}, so we include five GCLASS clusters at $z>1$ for much deeper follow up spectroscopy.  These clusters were themselves selected from SpARCS \citep{SpARCS09_Wilson,SpARCS09_Muzzin,Demarco_Sparcs}, a survey that identified clusters based on overdensities of "red-sequence" galaxies \citep[e.g.][]{GY00} using shallow $z^{\prime}$ and IRAC 3.6$\mu$m images over 42 square degrees.   In addition to the five GCLASS clusters we also include the next richest systems within the target redshift range; these are expected to be comparable to, or slightly less massive than, the GCLASS systems.   

To sample the most massive clusters, we include three spectroscopically confirmed clusters detected via their Sunyaev-Zeldovich (SZ) signature from the South Pole Telescope (SPT) survey.   Like the GCLASS sample, the SPT clusters have existing spectroscopy available on the brighter galaxies \citep{SPT0546,SPT2106,SPT0205}, so GOGREEN is primarily targeting the fainter objects.  

For the groups, we selected nine X-ray detected systems from the COSMOS and Subaru-XMM Deep Survey (SXDS) fields, in an analogous way to the selection made for GEEC2 \citep{GEEC2-survey}.  Deep, multiwavelength imaging and exquisite photometric redshifts already exist for these systems, enabling efficient targeting.  The COSMOS and SXDS groups are selected from updated versions of the catalogues described in \citet{alexis_sdf,alexis_cosmos} and \citet{George+11}. For target selection in COSMOS we use the UltraVISTA photometric catalogues of \citet{Ultravista_Muzzin}.  For SXDS we use an updated version of the UDS catalogues from \citet{W+09} and \citet{Quadri+12}, kindly provided by R. Quadri.  

The coordinates and redshifts of the 21 systems selected are given in Table~\ref{tab-cluster_sample}.  We select the targets to ensure they are reasonably distributed in redshift between $1.0 < z < 1.5$, and in RA and Dec for efficient observability from Gemini North and South.

\subsection{Multiwavelength Imaging}

We obtained deep $z^{\prime}$-band imaging on our twelve massive cluster targets, using GMOS-N (EEV) and GMOS-S (Hamamatsu) detectors, at the start of our program (end of 2014).  The nine group targets already have sufficiently deep $z^{\prime}$-band data for spectroscopic target selection from COSMOS and SXDS.  The GMOS observations are described in Table~\ref{tab-preim}.  GMOS-S observations, which were taken with the red-sensitive Hamamatsu detectors, were typically taken with integration times of 1.5 hours.  For the northern systems, integration times were typically 2.5 hours, to account for the lower sensitivity of the EEV detector.  There is some variation in these times to account for differences in observing conditions. Most systems were observed under 70 percentile seeing conditions ($\sim 0.7$\arcsec\ in $z^{\prime}$), 70 percentile cloud cover (up to $\sim 0.3$ mag extinction), and 80 percentile sky brightness.  A $3\times 3$ dither grid pattern was executed, with 6\arcsec\ steps. 

\subsubsection{GMOS $z^{\prime}$-band}\label{sec-zim} 
\begin{figure}
	{
		\includegraphics[clip=true,trim=0mm 0mm 0mm 0mm,width=3.in,angle=0]{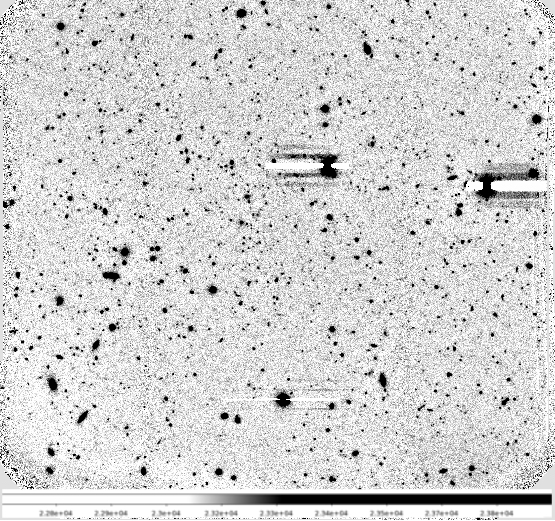}
	}
	\caption{The GMOS-S $z^{\prime}$-band image of SpARCS0035, highlighting the variable background across an amplifier in the presence of saturated stars.  The problem was fixed when the video board was replaced in May 2015.
		\label{fig-sparcs0035_zband}
	}
\end{figure}

The GMOS $z^{\prime}$-band data were reduced using the Gemini IRAF packages and standard procedures, including fringe correction.   Before May 2015, saturated pixels on the detector would affect the background level along the entire row of that amplifier.  An example is shown in Figure~\ref{fig-sparcs0035_zband}, for SpARCS0035.  This is primarily a cosmetic nuisance, but does eliminate a small fraction of the detector area from spectroscopic follow up. 

Zeropoints for the imaging were determined by comparing with pre-existing, but shallower, z$^{\prime}$ imaging from SpARCS (CFHT/MegaCAM) and the SPT collaboration (CTIO/MOSAIC-II).   The SpARCS zeropoints were obtained from standard stars taken during nighttime observations in the CFHT queue; the zeropoints are applied during the initial reduction stages by TERAPIX. We note that the GMOS z$^{\prime}$-band (particularly for the GMOS-S Hamamatsu chips) has a different wavelength coverage than the z$^{\prime}$-band on most cameras.  This is because while the transmission of the z$^{\prime}$-band filter itself typically extends up to 1.3$\mu$m, the effective wavelength is set by the declining quantum efficiency of the chips being used.  For most cameras, the transmission at $\lambda$ $>$ 9500\AA\ is negligible.  Both the deep depletion EEV chips and the Hamamatsu chips used here are more red-sensitive than typical CCDs, extending past 10000\AA, and therefore the effective wavelength of the z$^{\prime}$-band is longer.   However, a direct comparison of $z^\prime$ magnitudes taken with Gemini-S Hamamatsu and CFHT shows no significant offset, relative to the photometric uncertainties, as a function of magnitude.  We therefore neglect any colour term in the photometry. 

Magnitude limits are determined from the rms of ten pixel (1$\farcs$6) blank-sky aperture measurements across the field.  Preliminary 5$\sigma$ limits determined this way are given in Table~\ref{tab-preim}.  

\begin{table*}
	\begin{tabular}{lllllllll}
		Cluster &u/U&g/B/V &r/R &i/I &$z^{\prime}/z$&Y/J1&J&$K_s$\\
		Target depth &&26.5 &26.5 &25.0&26.0 &24.5&24.0&23.5\\
		\hline
		SpARCS0035 &VIMOS& VIMOS  & VIMOS & VIMOS&GMOS/VIMOS  &Fourstar &HAWK-I&HAWK-I \\
		SPT205 &VIMOS& VIMOS   & VIMOS  & VIMOS& GMOS/VIMOS  &Fourstar&Fourstar&Fourstar\\
		SpARCS0219 &VIMOS& VIMOS   & VIMOS  & VIMOS& GMOS/VIMOS  &Fourstar &Fourstar&Fourstar\\
		SpARCS0335 &VIMOS& VIMOS   & VIMOS  & VIMOS& GMOS/VIMOS  &HAWK-I &Fourstar  &Fourstar \\
		SPT0546 &VIMOS& VIMOS   & VIMOS  & VIMOS& GMOS/VIMOS  &Fourstar &Fourstar&Fourstar\\
		SpARCS1033 && SC & SC & SC &GMOS/HSC&HSC &&WHT\\
		SpARCS1034 && SC & SC & SC &GMOS/HSC&HSC+WIRCAM &WIRCAM&WIRCAM\\
		SpARCS1051 &Megacam& SC & SC & SC &GMOS/HSC&HSC &&\\
		SpARCS1616 &Megacam& SC & SC & SC &GMOS/HSC$^\ast$&HSC$^\ast$ &&\\
		SpARCS1634 &Megacam& SC & SC & SC &GMOS/HSC$^\ast$&HSC$^\ast$  &WIRCAM&WIRCAM\\
		SpARCS1638 &Megacam& SC & SC & SC &GMOS/HSC$^\ast$&HSC$^\ast$ &&\\
		SPT2106 &VIMOS & VIMOS   & VIMOS  &VIMOS& GMOS/VIMOS  &Fourstar&Fourstar &HAWK-I \\
	\end{tabular}
	\caption{Deep imaging observations of the SpARCS and SPT clusters in our sample, as of semester 2017A.  Table entries indicate instruments from which imaging data have been obtained, or for which observations are scheduled.  These are GMOS ($z^{\prime}$), VIMOS ($UBVRIz$) on the VLT, Suprimecam (SC) and HyperSuprimeCam (HSC) on Subaru ($grizY$), Fourstar on Magellan ($J_1JK_s$), HAWK-I on VLT ($YJK_s$) and WIRCAM on CFHT ($YJK_s$).  Target depths are indicated in the column headings, and listed observations are expected to be close to those depths.  We do not list a target depth for $u/U$ as this (non-critical) band is more heterogeneous.  Imaging is mostly complete apart from the northern clusters (SpARCS 10h and 16h clusters) for which J and $K_s$ observations are still required.   All targets also have deep {\it Spitzer} IRAC imaging with depths of at least 23.1 at 3.6$\mu$m. Entries marked with a $^\ast$ have been scheduled as of the time of writing.\label{tab-imaging}}
\end{table*}
\begin{figure*}
{\includegraphics[clip=true,trim=0mm 0mm 0mm 0mm,width=3.7in,angle=0]{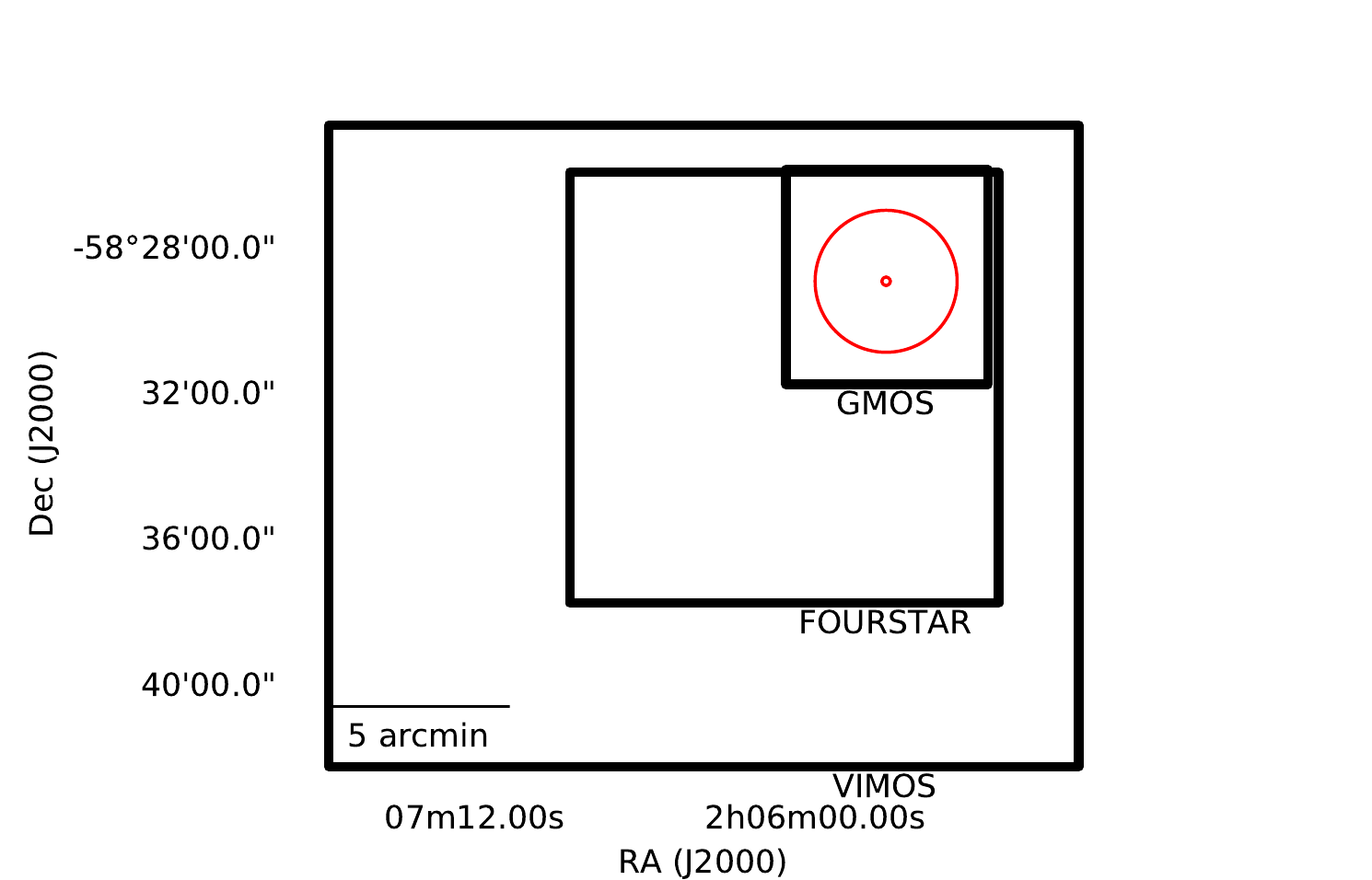}\includegraphics[clip=true,trim=0mm 0mm 0mm 0mm,width=3.7in,angle=0]{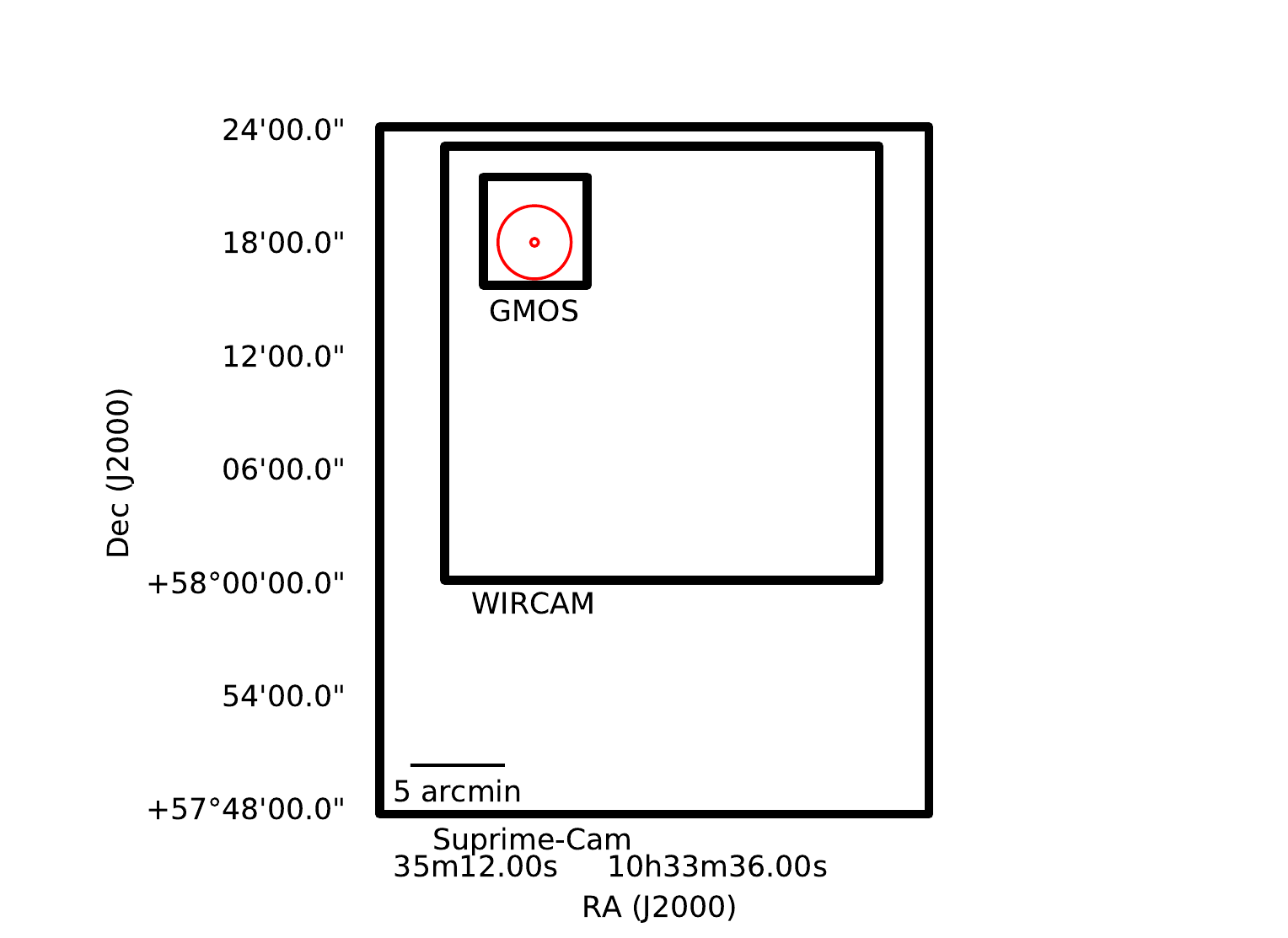}
	\caption{Examples of the imaging footprints for typical clusters, SPT-0205 (left panel) and SpARCS1034 (right panel). The smaller circle marks the cluster location and the larger circle is 1 Mpc (physical) at the cluster redshift, z=1.32 and 1.40 respectively. The rectangles (from smallest to largest) show the imaging fields of GMOS, FOURSTAR, and VIMOS (left) and GMOS, WIRCAM, and Suprime-Cam (right). For the four chip mosaic cameras (FOURSTAR, WIRCAM, HAWK-I and VIMOS), the cluster has been placed at the centre of one quadrant to ensure the full exposure depth is reached around the cluster centre in a single chip. The gaps in these mosaic cameras result in a region of lower exposure time in a central cross of the image.  
		\label{fig-footprint}}
}
\end{figure*}
\subsubsection{Spitzer IRAC imaging}
All but three of our clusters have publicly available deep ($5$-$\sigma$ depth of at least 2$\mu$Jy, or AB=23.1)  [3.6]$\mu$m imaging from {\it Spitzer} IRAC.  Most of the data
come from SERVS \citep{SERVS}, S-COSMOS \citep{COSMOS_Spitzer} and SpUDS 
\citep[PI J.  Dunlop, as described in][]{CANDELS}.  The three SPT clusters were observed as PI programs (PI Brodwin, from program ID 70053 and 60099).

The remaining three clusters only had imaging from SWIRE \citep{SWIRE}, which has a $5$-$\sigma$ depth of 7$\mu$Jy, sufficient only for the brighter targets in our sample.  We therefore obtained 1200s/pixel integrations over $3\times 2$ maps at [3.6]$\mu$m and [4.5]$\mu$m from Cycle 13 (PI McGee, GO program 13046).

\subsubsection{Other Optical and Near infrared imaging}\label{sec-mwi}
Multiwavelength imaging is required both to quantify the spectroscopic completeness and to determine the stellar masses and star formation histories of our galaxies.  In particular, broad wavelength coverage is crucial for classifying galaxies from their rest-frame colours.  We also require good photometric redshifts to understand the relevant completeness for cluster members \citep[e.g.][]{vdB+13}.  Well calibrated photometric redshifts will allow us to determine membership at radii outside of those probed by our GMOS spectroscopy.  

The nine group targets have existing deep, multiwavelength imaging spanning the full optical-NIR spectrum from COSMOS and SXDS, and our goal is to obtain comparably deep coverage in the same bands ($ugrizYJK$) for the other systems.  
\begin{figure*}
	{
		\includegraphics[clip=true,trim=0mm 0mm 0mm 0mm,width=7.5in,angle=0]{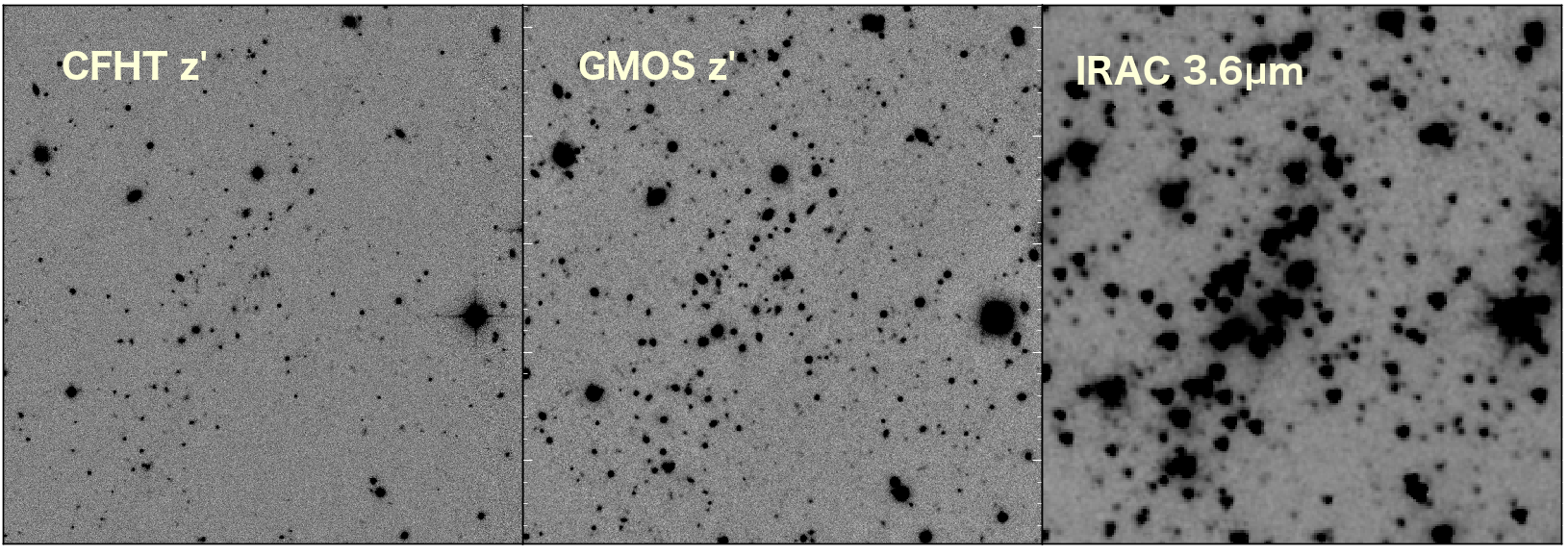}
	}
	\caption{An example of our imaging is shown, for SpARCS1634.  The left image shows the central 2.4\arcmin\ of the original SpARCS imaging available from CFHT MegaCam, from which the clusters were detected.  In the middle panel we present our new $z^{\prime}$ image from GMOS, and on the right we show the deep IRAC image. All images are oriented with North up and East to the left.
		\label{fig-imaging}
	}
\end{figure*}
To this end we have been using available resources to obtain homogeneous imaging on all systems.  The current status is described in Table~\ref{tab-imaging}.  Through observations on VLT, Magellan, Subaru and CFHT we expect to have obtained all required data except $JK_s$ for the northern systems, by the end of semester 17A.  Figure~\ref{fig-footprint} shows a typical field layout for two fields, SPT0205 in the south and SpARCS1034 in the north.

\subsection{Spectroscopy}
To obtain even low quality ($S/N \sim 2$ per \AA) spectroscopy on very faint ($z^{\prime}\sim 24$) galaxies requires exposure times of $\sim 15$ hours with GMOS.  To simultaneously achieve high completeness at brighter magnitudes, we observe each cluster with multiple slit masks, spread over several semesters.  Typically 25--30 slits can be assigned to priority targets on each mask, owing to geometrical constraints.  We allocate $\sim 15$ of the faintest galaxies ($23.5<z^{\prime} <24.25$) to every mask, such that they obtain 15h of total integration time.  Another $5$--$10$ slits per mask on brighter galaxies are different for each mask.  For massive clusters in which we have little or no existing data, we observe six masks of 3h each, to maximize the number of brighter targets.    By spreading the masks over three semesters we can make adjustments between observations; for example, by replacing faint targets that have reached the desired S/N prematurely. Most of the SpARCS clusters already have extensive spectroscopy from the GCLASS program; for these we plan only four masks of 5h each, focusing on the fainter galaxies.

The groups are significantly less rich, and there are fewer bright candidate members.  Therefore, we plan only three masks on each, with 5h exposures.  

\subsubsection{Spectroscopic selection catalogues}
Spectroscopic targets are selected directly from our deep z$^{\prime}$-band imaging, described in \S\ref{sec-zim}.  Target selection (see \S~\ref{sec-masks}) is made using simple magnitude and colour-cuts from combined z$^{\prime}$-band and IRAC 3.6$\micron$ photometry.  Figure~\ref{fig-imaging} shows an example of our deep imaging in these two bands for SpARCS1634, compared with the original CFHT image from which the cluster was detected in SpARCS.

Photometric catalogues for the spectroscopic selection were made following the methods laid out in \citet{Muzzin+08, SpARCS09_Muzzin} and \citet{SpARCS09_Wilson} for the SpARCS survey, and we refer the reader to those papers for full details.  In brief, for a given cluster, objects were detected separately in both the z$^{\prime}$-band and 3.6$\micron$ using the SExtractor package \citep{sextractor}.  Detection in separate filters has the advantage of being able to easily flag sources blended in the IRAC images, as well as being able to detect both extremely red and extremely blue objects not detected in complementary filters.  

Photometry in the z$^{\prime}$-band was performed in a fixed aperture of 3$\farcs$66 radius, which is chosen as a multiple of the IRAC native pixel scale (1$\farcs$22), and the SExtractor {\it mag\_auto} value was also recorded as the total z$^{\prime}$-band magnitude.  The 3.6$\mu$m filter photometry was performed in multiple fixed apertures ranging from 3$\farcs$66 -- 24$\farcs$0 radius.  The total IRAC magnitude was calculated using the method of \citet{Lacy05}, which effectively uses as the total magnitude the magnitude measured in the fixed aperture that is closest in size to the estimated isophotal radius of the galaxy as determined with SExtractor.  

Once objects are detected and fluxes measured in each band, objects are matched using a tolerance of 1$\farcs$0.  This is smaller than the FWHM of the $3.6\mu$m data, and therefore minimizes the number of spurious matches \citep[e.g.][]{Lacy05}.  \citet{Muzzin+08} estimated that this induces a spurious match rate of $\sim$ 4\%, most of which are caused by blended sources where the IRAC centroid is misplaced.  Therefore, we note that there may be catastrophic photometry for as many as 4\% of sources in the selection catalogues.  However, since it is caused by random blends primarily of foreground/background galaxies this should not bias the selection of spectroscopic GOGREEN targets.  We emphasize that these methods are only used for constructing catalogues for selecting spectroscopic targets.  Future multi-wavelength catalogues will use PSF matching and fitting techniques to mitigate the effect of blending \citep[e.g.][]{vdB+13}.

The z$^{\prime}$-$3.6\mu$m colours are measured using the 3$\farcs$66 radius apertures, with an aperture correction for the flux lost from the non-Gaussian wings of the IRAC PSF \citep{Lacy05}.  PSF homogenization is not done for the colours because degradation of the deep z$^{\prime}$-band image quality ($\sim$ 0$\farcs$7) to the poorer image quality of the IRAC data ($\sim$ 1$\farcs$8) would cause significant blending and affect the colour measurements.  The 3$\farcs$66 radius aperture is larger than the FWHM in both z$^{\prime}$-band and $3.6\mu$m, and larger than the typical size of high-redshift galaxies ($\sim$ 1$\farcs$0) and therefore provides an unbiased colour without the need for PSF homogenization, at the sacrifice of some signal-to-noise.  This method for photometry was used extensively in SpARCS \citep[e.g.][]{SpARCS09_Muzzin} and other wide-field Spitzer surveys \citep[e.g.][]{IRACSS} and has been shown to provide reliable colours.  
For all clusters a clear red-sequence is visible (see \S\S~\ref{sec-masks} and \ref{sec-status}), which gives confidence that the photometry is of sufficient quality to select spectroscopic targets.

Stars are identified in the z$^{\prime}$-band using the SExtractor {\it class\_star} parameter.  This is important both for marking potential mask alignment stars and telluric standards, and for avoiding selecting stars as science targets.  

 \begin{figure}{
 		\includegraphics[clip=true,trim=0mm 0mm 0mm 0mm,width=3.5in,angle=0]{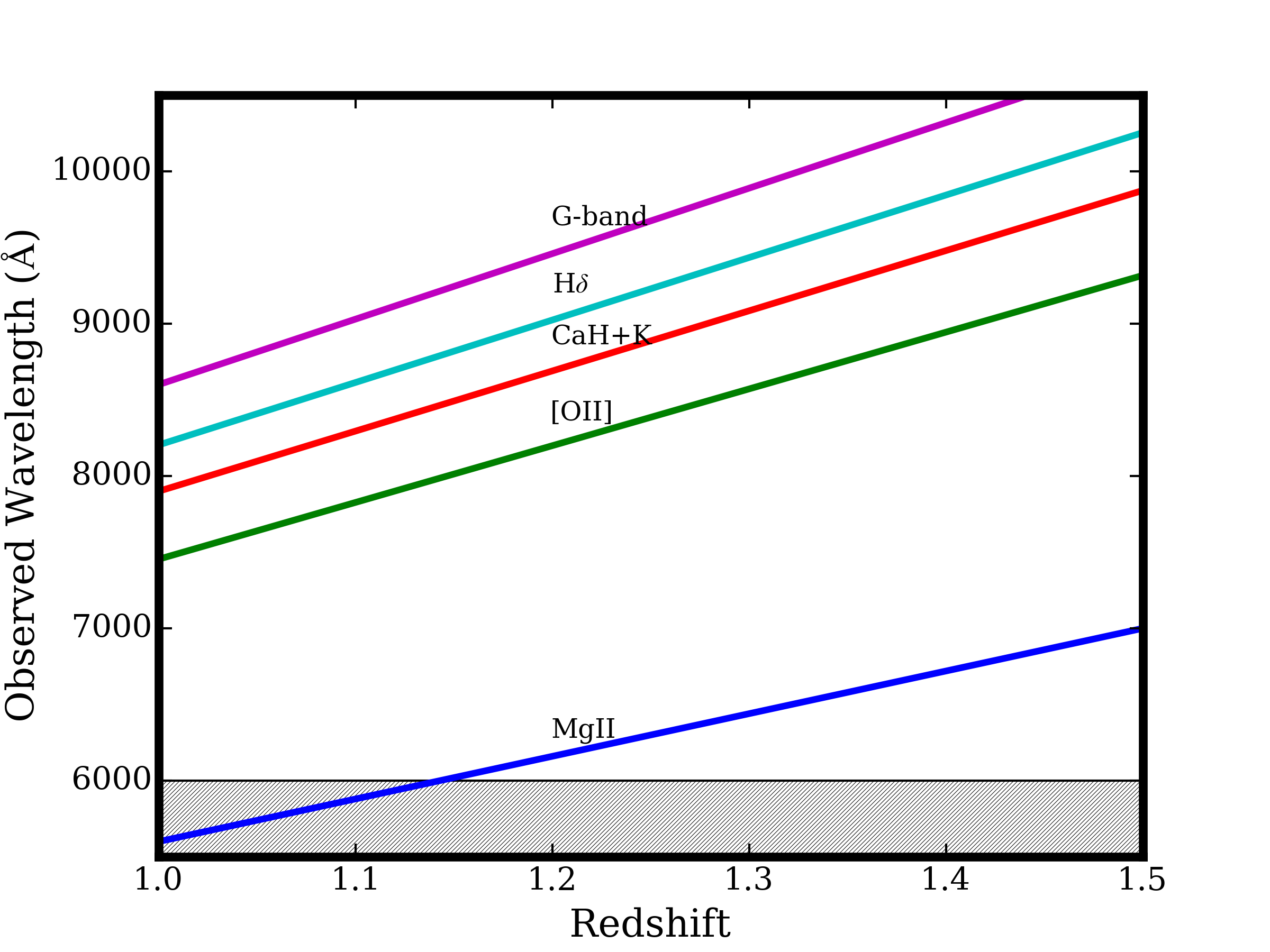}}
 	\caption{The coverage of key spectral features is shown as a function of redshift, over the target redshift range $0<z<1.5$.  While good data are obtained over the full wavelength range $5500$--$10500$\AA, the wavelength calibration is unreliable below $6000$\AA, indicated as the shaded region.  With the good red sensitivity and accurate sky subtraction, we are able to identify the usual UV and optical line indices over the full redshift range.\label{fig-speccoverage}}
 \end{figure}

\subsubsection{Instrument Configuration}
Spectroscopy is obtained with the GMOS-S and GMOS-N instruments, which cover a 5.5$\times$5.5\arcmin\ field of view.  All observations on GMOS-S were obtained with the Hamamatsu detector array, which consists of three chips.  Two of these have enhanced red response, while the chip at the blue end has enhanced blue response.  Pixels are 15$\mu$m on a side, corresponding to 0.080 arcsec/pixel.  All our observations are obtained with the detector binned 2$\times$2, resulting in a pixel scale of $0\farcs16$.  On GMOS-N, observations prior to 2017 were obtained with an array of identical EEV deep depletion detectors.  These detectors have a pixel scale of $0\farcs0727$; as with the GMOS-S data we bin $\times$2 for a final pixel scale of $0\farcs145$.  In 2017 the GMOS-N detector was replaced with a Hamamatsu array identical to the one on GMOS-S.

We observe all fields with the R150 grating, in nod and shuffle mode.  The low resolution is chosen to maximize the wavelength coverage on the detector, ensuring that redshift completeness is high.  With the $2\times2$ detector binning, the dispersion is 3.9\AA\ for the Hamamatsu detectors (GMOS-S), and 3.5\AA\ for the EEV (GMOS-N).  Slits are 1\arcsec\ wide, resulting in a spectral resolution of $\sim 460$, or $\sim 20$\AA.  

The slits are 3\arcsec\ long, and we centre the object $0\farcs725$ away from the centre of the slit.  The telescope is then nodded by $1\farcs45$, placing the object $0\farcs725$ on the other side of the centre.  Most of our masks are observed in microshuffle mode, where charge is shuffled by a little more than a slit width.  For most clusters we also observe a mask in band-shuffle mode, where the charge is shuffled by a third of the detector height.  This is done to achieve maximum target density in the core of the rich clusters.

Figure~\ref{fig-speccoverage} shows how our wide wavelength range enables coverage of key spectral features, from the MgII absorption line at $2800$\AA\ to the G-band at 4300\AA, over the full redshift range $1.0<z<1.5$. 

We use spectral dithers, observing each mask at three different central grating settings (8300, 8500, and 8700 \AA).  This allows contiguous wavelength coverage in the presence of chip gaps and bad columns on the detector.
\begin{figure}
	{
		\includegraphics[clip=true,trim=0mm 0mm 0mm 0mm,width=3.6in,angle=0]{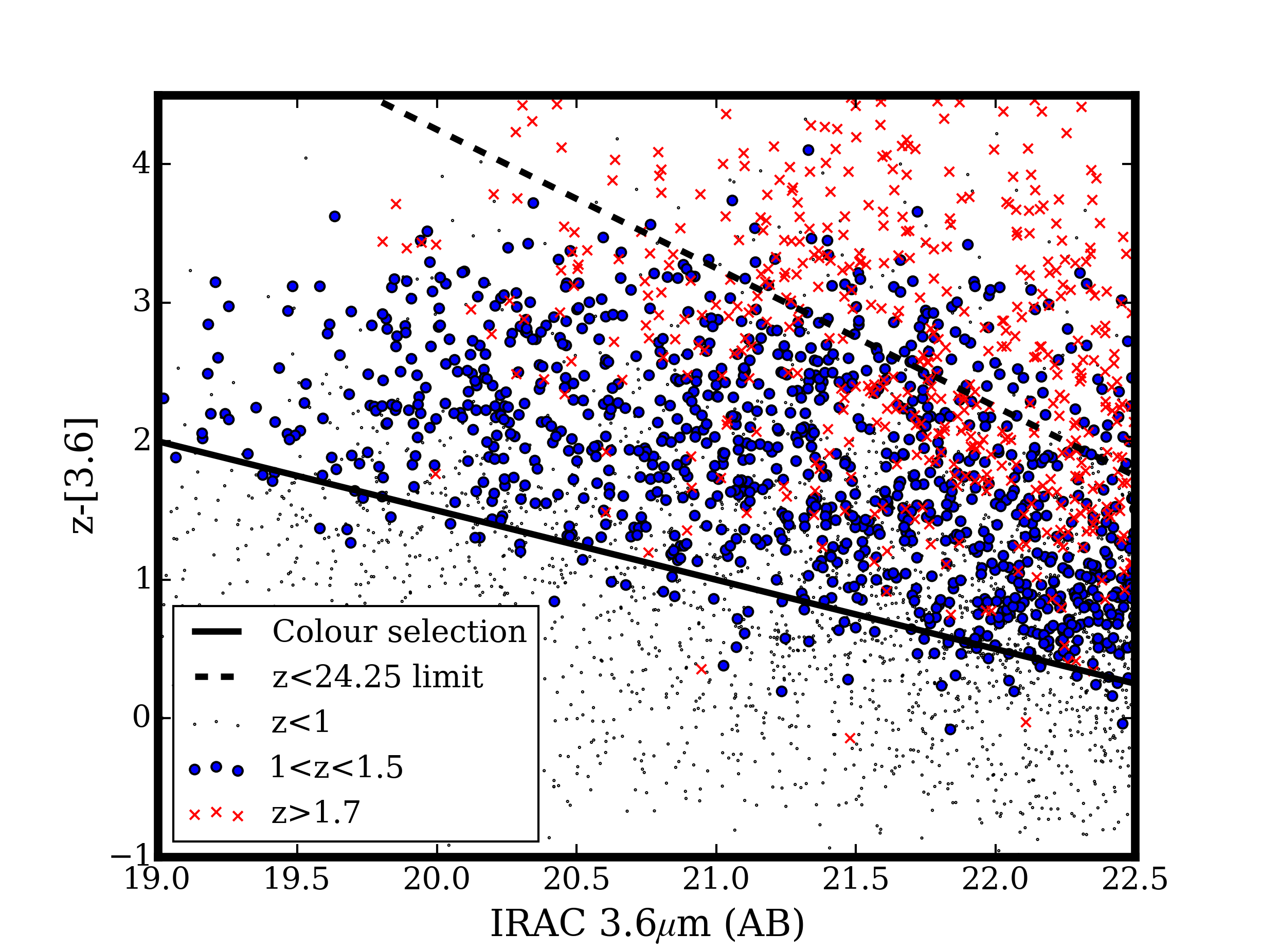}
		
	}
	\caption{(z-[3.6]) as a function of [3.6] magnitude for a random subset of UltraVISTA data \citep{Ultravista_Muzzin}, coloured by photometric redshift.  Targets are selected to be redder than the solid black line, to remove contamination from $z<1$ galaxies.  The $z^{\prime}-$band limit of the survey, $z^{\prime}<24.25$, is shown by the dashed line and naturally excludes most high redshift galaxies.  Galaxies with $z>1.7$ are shown by red crosses, and will generally not get a measured redshift because [OII]$\lambda$3727 is shifted out of our wavelength range.
		\label{fig-UVCMD}
	}
\end{figure}

\subsubsection{Spectroscopic Target Selection and Mask Design}\label{sec-masks}
Galaxy targets are selected based on their $3.6 \mu$m and $z^{\prime}-$band flux from deep IRAC and GMOS imaging. Specifically, galaxies must have total magnitudes $[3.6]<22.5$ and $z^{\prime}<24.25$.  To target this range efficiently requires some colour preselection to remove foreground and background galaxies.  We take different approaches for the 12 clusters (SpARCS and SPT systems), which are fairly rich but lack photometric redshifts, and for the nine groups which are poor but have the advantage of exquisite photometric redshifts.

\begin{figure}
	{
		\includegraphics[clip=true,trim=0mm 0mm 0mm 0mm,width=3.7in,angle=0]{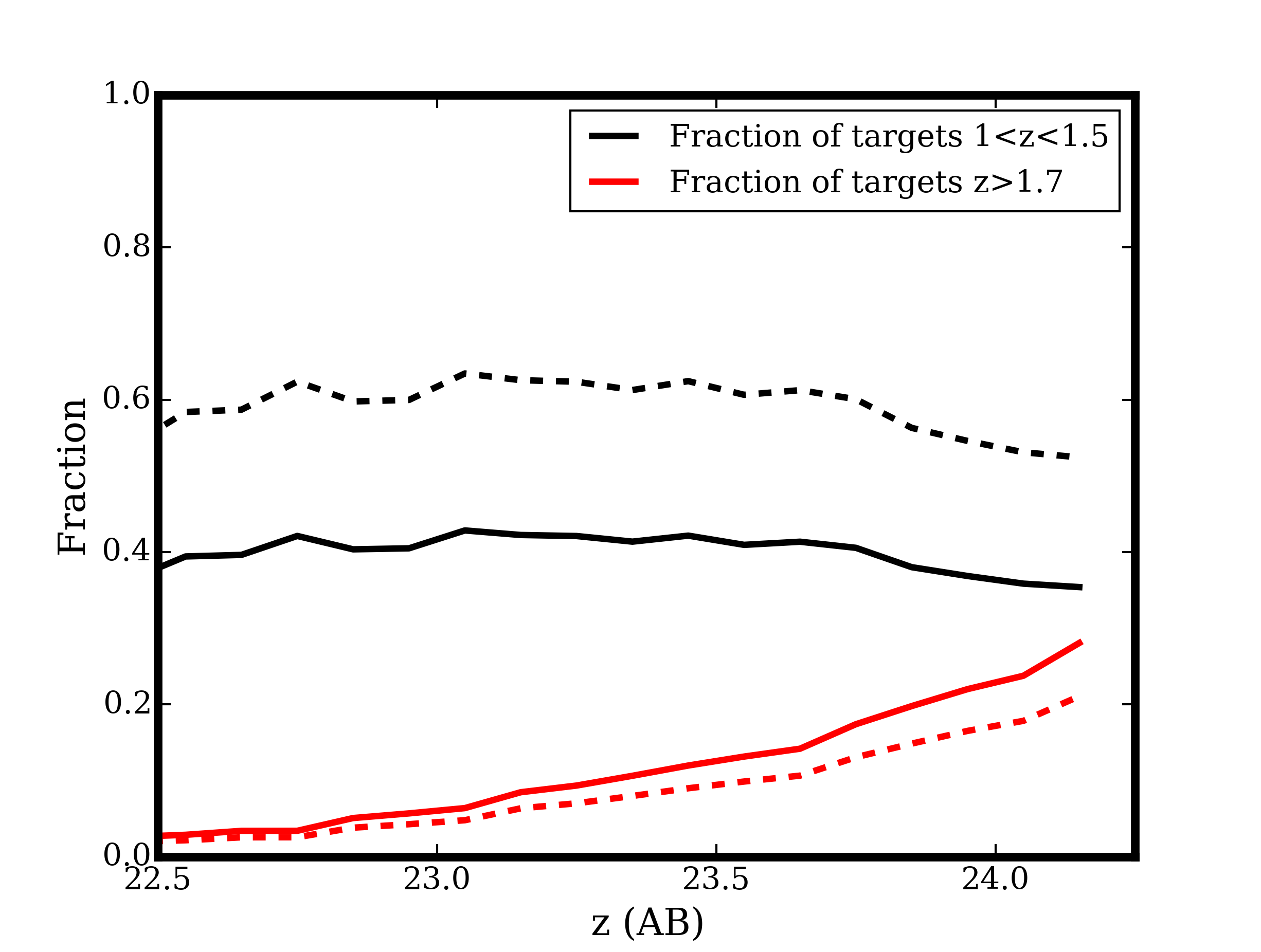}
		\caption{The expected success of our colour selection is shown as a function of total $z^{\prime}$ magnitude.  The solid black line shows the fraction of targeted galaxies that are expected to lie in the redshift range $1.0<z<1.5$, based on the UltraVISTA \citep{Ultravista_Muzzin} photometric redshift sample.  The red line shows the fraction of targeted galaxies expected to lie at $z>1.7$, for which it is unlikely that we would be able to obtain a redshift.  These fractions are derived from a field average, while our targeted areas have massive clusters in the target redshift range.  If we assume the $1.0<z<1.5$ slice distributed over the GMOS field of view is moderately overdense by a factor $\sim 2$,  we obtain the dashed lines.  Thus we expect about 50\% of our targets to lie in the required redshift range, with only $\sim$20\% high redshift contamination at the faintest magnitudes.\label{fig-fsample}}
	}
\end{figure}

For the cluster samples we use a colour cut for avoiding low redshift ($z<1$) contamination.  This is determined by examining the colour-magnitude distribution of galaxies in UltraVISTA with good photometric redshifts \citep{Ultravista_Muzzin}.  A random subset of this sample is shown in Figure~\ref{fig-UVCMD}, with different symbols representing galaxies at $z<1$, $1.0<z<1.5$ and $z>1.7$. The latter will have the [OII] emission line redshifted beyond our wavelength range and thus we are unlikely to be able to measure a redshift.  A simple colour cut  $(z^{\prime}-[3.6])>2-0.5([3.6]-19)$ is made to exclude low redshift galaxies.  The effectiveness of these cuts is shown in Figure~\ref{fig-fsample}.  The solid black line shows the expected fraction of all primary targets that lie in the desired redshift range $1.0<z<1.5$ in an average patch of UltraVISTA.  This is $\sim 40$ per cent, roughly independent of magnitude. However, our target fields are not average patches, but host massive clusters; thus our success rate is expected to be significantly higher than that.  The dashed line shows the result if the field is overdense in the $1.0<z<1.5$ redshift slice by a modest factor of two, and this raises the efficiency to $\sim 60$ per cent.  The red lines show the fraction of $z>1.7$ galaxies that will be targeted; this rises to at most $\sim 30$ per cent at the faintest magnitudes, and only $\sim 20$ per cent in the presence of an overdense region.  The blue colour cut shown in Figure~\ref{fig-fsample} excludes about 7 per cent of $1.0<z<1.5$  galaxies at $z^{\prime}<24.25$ and $[3.6]<22.5$.
 
\begin{figure}
{
	\includegraphics[clip=true,trim=0mm 0mm 0mm 0mm,width=3.7in,angle=0]{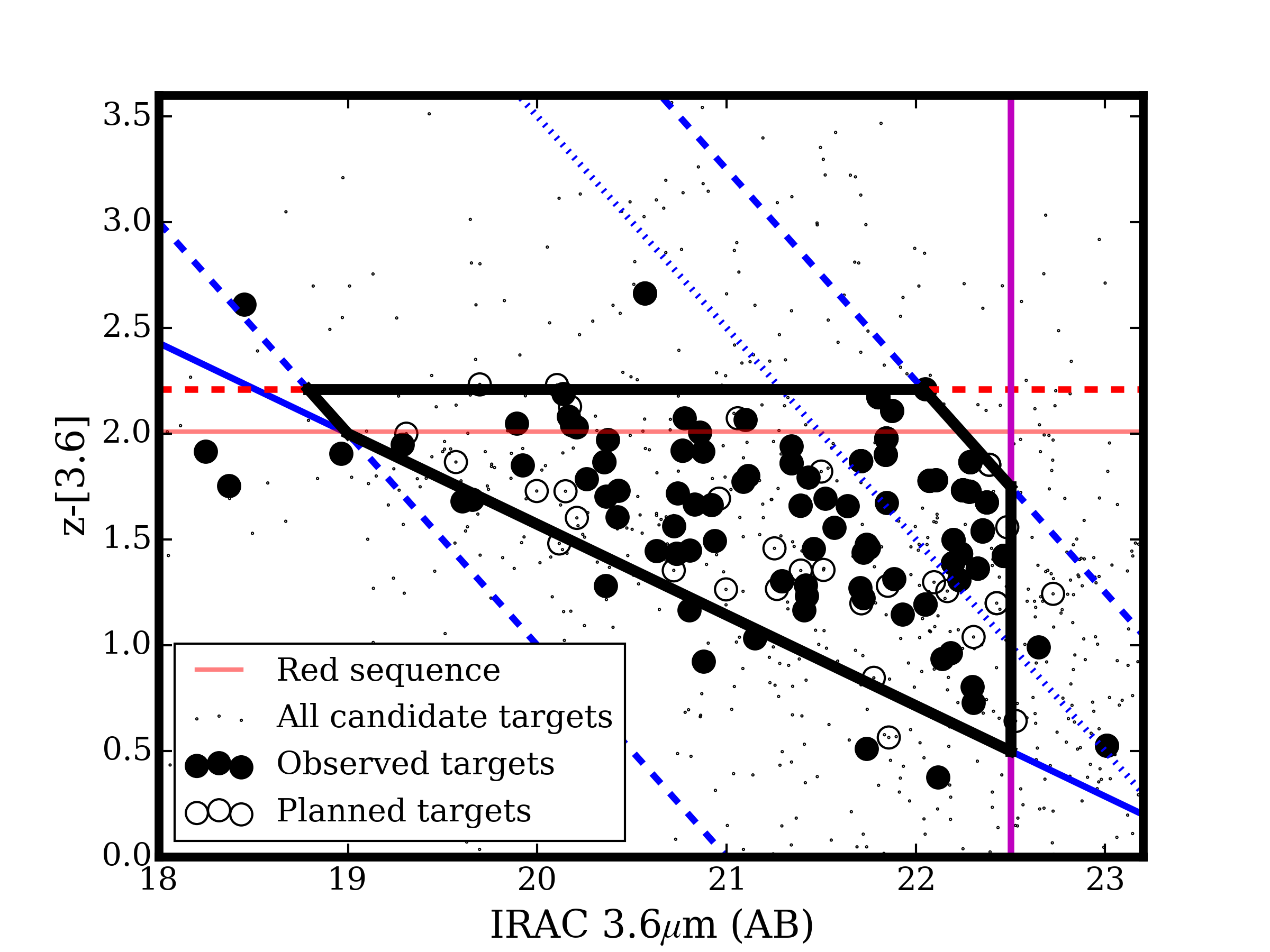}
	\caption{The colour-magnitude diagram of SPT0546 is shown, with all galaxies detected in $z^{\prime}$ and $IRAC$ within the GMOS field of view shown as small dots.  The thick black lines outline the selection area for our primary target sample.  This is bounded by the IRAC limit of 22.5 (solid, vertical magenta line), the $21<z^{\prime}<24.5$ limits (thick, blue dashed lines), a colour cut to exclude foreground galaxies (thick blue solid line) and a cut 0.2 mag brighter than the red sequence (dashed red line, with the red sequence itself shown as the solid, red line).  The dotted blue line indicates $z^{\prime}=23.5$; primary targets fainter than this are observed on multiple masks to increase exposure time.  Six masks were designed for this cluster, and four have been observed. Large, filled points indicate galaxies already observed, while large open symbols are those allocated to masks that have not yet been observed. Some targets that lie outside the colour selection boundaries are included in the masks as ``fillers'', once the mask is fully populated with priority targets. \label{fig-SPT0546mask}}
}
\end{figure}

\begin{table*}
	\begin{tabular}{lllllll}
		Target &Date &Mask &Band/Micro &Telescope/ &Integration &Notes\\
		&       &     &           &Detector   & time (ks)  &           \\
		\hline
		SPT0205 & Nov 16,18, 2014 & GS2014BLP001-06 & Microshuffle & GS/Ham & 6.48&\\ 
		& Oct 29-30, Nov 3 2016& GS2016BLP001-02 & Microshuffle & GS/Ham &10.8 &\\ 
		& Oct 28-29, 2016	& GS2016BLP001-09 & Microshuffle & GS/Ham & 9.36&\\ 
		\hline
		SPT0546 & Nov 15-16, 2014 & GS2014BLP001-09 & Microshuffle & GS/Ham & 5.76&\\ 
		& Nov 17,19, 2014 & GS2014BLP001-10 & Microshuffle & GS/Ham & 7.2&\\ 
		& Nov 20, 2015 & GS2015BLP001-15 & Microshuffle & GS/Ham & 7.92 & \\ 
		& Nov 21, 2015 & GS2015BLP001-16 & Microshuffle & GS/Ham & 2.16 & \\ 
		& Feb 10, 2016 & GS2015BLP001-16 & Microshuffle & GS/Ham & 14.4 & \\
\hline
		SpARCS0035 & Nov 21, 2015 & GS2015BLP001-05 & Bandshuffle & GS/Ham & 9.36 & \\
		& Nov 20, 2015 & GS2015BLP001-06 & Microshuffle & GS/Ham & 7.2 & \\ 
		& Oct 28, 2016 & GS2016BLP001-01 & Microshuffle & GS/Ham & 7.9& \\
		& Oct 27, 2016 & GS2016BLP001-07 & Microshuffle & GS/Ham & 10.8& \\ 
\hline
		SpARCS0219 & Nov 20, 2015 & GS2015BLP001-17 & Microshuffle & GS/Ham & 10.8 & \\ 
		& Oct 30, 2016& GS2016BLP001-03 & Microshuffle & GS/Ham &8.64 & \\ 
		& Oct 27-28, 2016	& GS2016BLP001-12 & Microshuffle & GS/Ham & 9.36& \\ 
\hline
		SpARCS0335 & Nov 18-19, 2014 & GS2014BLP001-01 & Bandshuffle & GS/Ham & 7.2&\\ 
		& Feb 1, 2017	& GS2016BLP001-13 & Bandshuffle & GS/Ham & 9.36&\\ 
		& Oct 26-29, 2016	& GS2016BLP001-14 & Bandshuffle & GS/Ham &10.8 &\\ 
		\hline
		SpARCS1051 & Feb 18\&29, 2016 & GN2016ALP004-03 & Microshuffle & GN/EEV & 18.0 & Queue\\
				   & April 25, 2017 & GN2017ALP004-08 & Microshuffle & GN/Ham & 12.0&\\
				   & April 26, 2017 & GN2017ALP004-07 & Microshuffle & GN/Ham & 13.8&\\
				   \hline
		SpARCS1033 & April 18, 2017 & GN2017ALP004-01 & Bandshuffle & GN/Ham & 7.2&\\
		& April 19, 2017 & GN2017ALP004-02 & Microshuffle & GN/Ham & 10.08&\\
		& April 20, 2017 & GN2017ALP004-03 & Microshuffle & GN/Ham & 10.08&\\
		\hline
		SpARCS1034 & April 24, 2017 & GN2017ALP004-04 & Bandshuffle & GN/Ham & 4.3&\\
					& April 12\&27, 2017 & GN2016ALP004-05 & Bandshuffle & GN/Ham & 10.08 & \\
					\hline
		SpARCS1616 & June 1, 2016 & GN2016ALP004-06 & Microshuffle & GN/EEV & 14.4 & \\
				   & June 2, 2016 & GN2016ALP004-07 & Microshuffle & GN/EEV & 18.0 & \\
				   & April 18\&27, 2017 & GN2016ALP004-09 & Microshuffle & GN/Ham & 17.28 & \\
				   \hline
		SpARCS1634 & May 30, 2016 & GN2016ALP004-04 & Microshuffle & GN/EEV & 10.8 & \\
				   & May 30-31, 2016 & GN2016ALP004-05 & Microshuffle & GN/EEV & 18.0 & \\
				   & April 19/26, 2017 & GN2016ALP004-10 & Microshuffle & GN/Ham & 18.0 & \\
				   \hline
		SpARCS1638 & May 28-20, 2016 & GN2016ALP004-01 & Microshuffle & GN/EEV & 10.8 & \\
					& May 29/June2, 2016 & GN2016ALP004-02 & Microshuffle & GN/EEV & 18.0 & \\
					& April 20/28, 2017 & GN2017ALP004-11 & Microshuffle & GN/Ham & 18.0 & \\
					\hline
		COSMOS-28 & Jan 30, 2016 & GN2015BLP004-03 & Microshuffle & GN/EEV & 18.0&Queue \\ 
		\hline
		COSMOS-63 & Jan 31, 2016 & GN2015BLP004-02 & Microshuffle & GN/EEV & 18.0&Queue\\ 
		\hline
		COSMOS-125 & Jan 31, 2016 & GS2016ALP001-02 & Microshuffle & GS/Ham & 15.12 &\\
		& Feb 25, 2015 & GS2015ALP001-02 & Microshuffle & GS/Ham & 12.25&\\ 
		\hline
		COSMOS-221 & Feb 24, 2015 &  GS2015ALP001-01 & Microshuffle & GS/Ham & 10.08&\\ 
		& Feb 23, 2015 & GS2014BLP001-05 & Microshuffle & GS/Ham & 5.04&\\ 
		& Feb 13, 2016 & GS2016ALP001-01 & Microshuffle & GS/Ham & 10.8 & \\
		\hline
		SXDF49 & Oct 9, 2015 & GN2015BLP004-01 & Microshuffle & GN/EEV & 18.0&Queue\\ 
		\hline
		SXDF64 & Nov 17, 2014 & GS2014BLP001-08 & Microshuffle & GS/Ham & 7.2&\\
		\hline
		SXDF76 & Nov 15, 2014 & GS2014BLP001-02 & Microshuffle & GS/Ham & 5.76&\\
		\hline
		SXDF87 & Nov 15, 2014 & GS2014BLP001-07 & Microshuffle & GS/Ham & 8.64&\\
		\hline
	\end{tabular}
	\caption{A log of all spectroscopic data obtained as of April 2017 (mid-semester 2017A).  All data were acquired in Priority Visitor mode unless otherwise indicated in the Final column.  \label{tab-specobs}}
\end{table*}

For the mask design, in addition to the broad cuts described above, we fit the red sequence in $z^{\prime}-[3.6]$ colour, with a slope of zero.  An initial estimate of the colour is made based on the redshift of the cluster and the models of \citet{BC03} as described in \citet{SpARCS09_Muzzin}.  When necessary, this is adjusted based on the overdensity of galaxies on the colour-magnitude diagram.  The adopted colours are given in Table~\ref{tab-cluster_sample}.   Only galaxies up to 0.2 mag redder than this red sequence are considered primary targets.

In order to optimize the mask design, we then use a Monte Carlo technique, whereby the complete set (3--6 masks) is designed together, and 1000 realisations of each set is performed. The overall aim of the design is to obtain high numbers of galaxies in the bright ($z<23.5)$ and faint ($z^{\prime}>23.5$) bins, and ensure reasonable completeness in the cluster core where geometry maximally constrains slit placement and would otherwise lead to underrepresented galaxies simply due to slit collisions. In order to do this we use the following figure of merit (FOM) to evaluate the mask designs:
\begin{equation}
\mbox{FOM}=
\begin{cases}
0.5n_{\rm f} +0.05n_{\rm b},  & {\rm if}\, n_{\rm f} < 11 \\
0.2n_{\rm f} +1.0n_{\rm b}, & {\rm otherwise}
\end{cases}
\label{eqn:merit}
\end{equation}
where $n_{\rm b}$ and $n_{\rm f}$ represent the number of bright and faint objects, respectively, within 1Mpc of the cluster core allocated slits.  This naturally downweights masks where there are insufficient faint galaxies in the final mask (r$\leq$1 Mpc). These are the most difficult to allocate as they must be allocated on every mask, and so negating this step naturally favours bright galaxies which need only be allocated to a single mask. The set of masks with the highest score from the 1000 realisations is used. As mentioned above, for cluster targets, one of the masks is typically band-shuffled, and so covers the central third of the GMOS field but with higher target density.  This mask only contains bright galaxies, which would otherwise be underrepresented in the final sample due to geometry constraints. An example of the final target selection in colour-magnitude space for one of the clusters, SPT0546, is shown in Figure~\ref{fig-SPT0546mask}.   
The figure shows the total sample from six GMOS masks, with galaxies already observed indicated with filled points.

For the massive group sample, the exquisite deep optical and X--ray data in COSMOS, CDFS and SXDS make it possible to perform similar analysis on much lower mass haloes, following the GEEC2 strategy \citep{GEEC2-1}.  In particular the high-precision photometric-redshifts available improve target selection efficiency to a level comparable to that of the colour--selected cluster fields, without introducing significant bias.  Instead of a straight sum of the number of galaxies, a weight $W$ is applied based on each galaxy's photometric redshift ($z_{\rm ph}$), uncertainty ($\sigma_{\rm zph}$), and its relation to the cluster redshift ($z_{\rm clus}$)

\begin{equation}
W = 
\begin{cases}
1, & {\rm if}\, |z_{\rm ph} - z_{\rm clus}| \leq 2\sigma_{\rm zph} \\
0.5 & {\rm if}\, 0.7 \leq z_{\rm ph} \leq 2.0 \\
0.0 & {\rm otherwise}. 
\end{cases}
\end{equation}
These weights are summed as in Equation~\ref{eqn:merit} to determine the best mask set. 

Once the optimal set of primary target masks has been designed, the masks are examined to see where space for additional slits exists, and filler targets outside the primary sample are added.  These extra targets are drawn from all galaxies in the merged $z^{\prime}$ and IRAC catalogue.

\section{Spectroscopic Observations and Data Reduction}\label{sec-obs}
\begin{figure*}
	{
		\includegraphics[clip=true,trim=0mm 0mm 0mm 0mm,width=7.in,angle=0]{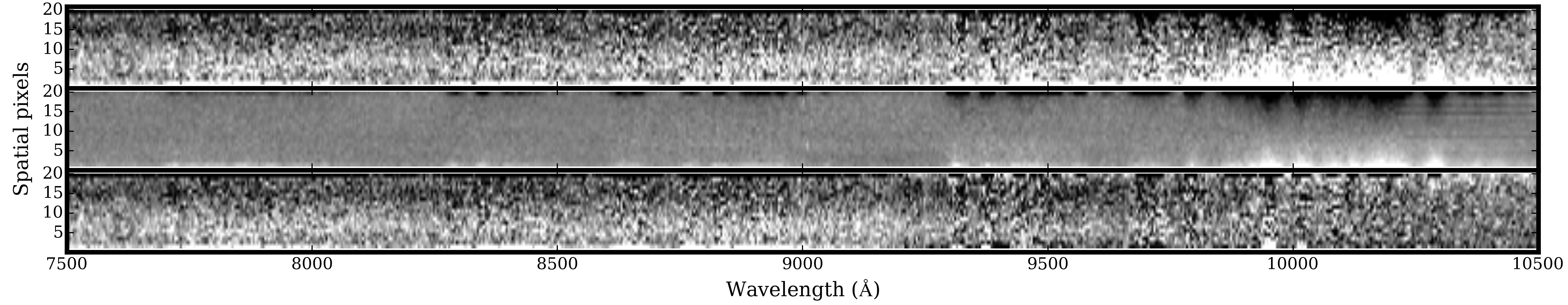}
	}
	\caption{
		The top image shows the red end of a sky-subtracted, two dimensional spectrum from a single slit in mask GS205ALP001-02.  Strong residuals are evident at $\lambda>9500$\AA\ as positive flux near the bottom of the frame, and negative flux near the top.  This is due to charge diffusion from the nod-and-shuffle pair, as described in the text.  To correct this, we create a template from a stack of sky-subtracted spectra, with continuum removed.  This is shown in the middle panel, where the charge diffusion residuals are the only feature.  After applying this correction to the data, we obtain the spectrum in the bottom panel.  The greyscale is the same in all three images, ranging from $-10$ to $+10$ counts.
		\label{fig-residuals}
	}
\end{figure*}

The spectroscopic data reduction is based on the {\sc iraf} tools provided by Gemini, via the {\it Ureka} distribution.  A variance and data quality (DQ) plane are propagated through all the reduction steps.
Table~\ref{tab-specobs} describes when the spectroscopic data were acquired, as of mid-semester 2017A.  Forty-six independent masks have been observed, of which seven are in band-shuffle configuration.  The weather condition constraints for this program are CC70 ($<0.3$mag extinction) and IQ70 (0.5\arcsec--0.7\arcsec in $z^{\prime}$ at zenith).  However, most of these observations were carried out in Priority Visitor mode, where the visiting observers are able to choose when to execute their program, within an observing run that is generally longer than the allocated time.  In many cases this allowed us to take advantage of better conditions than planned for.  

\subsection{Detrending and Sky subtraction}
Bias frames observed close to each observation are combined and subtracted from all data.  Bad pixel masks are created from the masks provided in the {\sc iraf} distribution, with additional bad pixels and columns identified from dark frames.  Dark frames were observed in every semester, and these are also subtracted from the data.  In semester 2016B, structure at the level of several hundred counts appeared in the GMOS-S detectors.  This is correctable with bias subtraction, but the structure is variable from night to night.  For these data, a unique bias frame is generated for every science frame, by linearly interpolating the two bias frames that bracket the science data, based on the time of observation.  

Flat field frames are interspersed with science frames, to allow accurate slit identification.  The data are not flat field corrected, however, as  the statistical noise introduced by flat fielding is generally larger than any systematic effect it corrects. 
Cosmic ray rejection is performed using {\it gemcrspec}, which is a wrapper for the LA Cosmic routine \citep{lacosmic}.

The GMOS-S detector has three CCDs, each with a different QE as a function of wavelength.  This is corrected using the {\it gqecorr} routine provided by Gemini, which generates a wavelength-dependent correction given a wavelength calibrated frame (in our case an arc) and a flat field frame.  This correction is then applied to the wavelength-calibrated, sky-subtracted science frame.
All our science data are taken in nod-and-shuffle mode.  Thus, sky subtraction is done simply by subtracting the science image from the corresponding sky image.  We also produce a ``sky'' spectrum by {\it adding} the two images.  This is useful for checking the wavelength calibration (see below) and for distinguishing sky residuals from emission lines in our science data.

\subsection{Wavelength Calibration}
Wavelength calibration is done using CuAr arc lamps, usually taken after a night's observing.  At our low resolution, this lamp provides $\sim 10$ useful lines over the wavelength range $6200\mbox{\AA}<\lambda<10700$ \AA.  The typical {\it rms} of the wavelength solution is $\sim 0.5$\AA.
All spectra (from both GMOS-S and GMOS-N) are linearized and rebinned to 3.91\AA\ per pixel, and forced to span $5500\mbox{\AA}<\lambda<10500$\AA.  
In general the wavelength calibration is not robust for $\lambda\lesssim6000$\AA, due to the lack of good arc lines at this resolution. 

To account for simple shifts in the zeropoint due to instrument flexure, we cross-correlate each sky spectrum with that of a reference slit, ideally chosen to have an accurate wavelength solution.  The median shift for each mask is computed, and applied to the wavelength solution of that mask.  Shifts are typically $<0.5$ pixels, though on occasion can be two or three times larger.  

The final wavelength calibration is applied to the ``sky'' spectra described above; all slits in a mask are then aligned in wavelength and displayed for a careful visual check of the wavelength solution.  

\subsection{Charge Diffusion Correction}\label{sec-cdc}
Charge on the detector diffuses away from its original pixel, by a distance that increases with wavelength.  This effect was described by \citet{GDDS}.  Because of the wavelength dependence, it is even more of a concern when using the Hamamatsu detectors, which have significant sensitivity beyond 1 $\mu$m.  In our data, charge from bright sky lines spreads as far as 10 binned pixels, or $1\farcs6$.  This is a serious problem in microshuffle mode, where the charge from the two nodded positions is typically separated by only one or two pixels.  This results in sky residuals that do not subtract, in every slit.  An example is shown in the top spectrum of Figure~\ref{fig-residuals}.  The same effect will also cause residuals in neighbouring slits when placed close together; however as this is much more difficult to correct for, and affects $\lesssim 10$ per cent of slits, we neglect it for now.  

Because of our large data volume, we are able to implement an empirical correction that works well for most of the masks.  First we combine a set of sky-subtracted, wavelength calibrated two-dimensional science spectra.  All spectra must have the same shuffle distance, which was either 38 or 40 pixels for the GMOS-S data.  We need to consider each science slit, as well as its associated sky slit (produced by adding, rather than subtracting, data from the two nod positions).   Slits with mean counts $>30$ within the wavelength range $9000\mbox{\AA}<\lambda<9250$\AA\ are excluded; this was determined empirically as necessary to exclude some bad slits.  Finally, we exclude data with $57000<\mbox{MJD}<57100$ and $x_{ccd}<1000$ because, as we discuss below, these data are affected by an additional contribution.  The selected slits are then combined with a weighted average, masking pixels with either a DQ flag, NaNs in either the science or sky frame, values of $<0$ in the sky frame or absolute values $>100$ in the science frame.  The weights are the inverse of the mean sky counts within the wavelength range $9800\mbox{\AA}<\lambda<10000$\AA; this is a region of bright sky emission lines that produce the most detrimental effect on our science data.   This produces a clean, high-S/N average of all our spectra; it includes the average science signal as well as any residuals not removed from nod-and-shuffle sky subtraction.  We also average the corresponding sky spectra, in exactly the same way. 

The next step is to remove the average science signal from the average spectrum above.  We do this by constructing an average of the sky-subtracted, wavelength-calibrated data from all band-shuffle slits.  Since bandshuffle spectra are well separated on the detector, slit pairs are not contaminated by the charge diffusion, and the stack yields a 2D spectrum of the average science data, free from residuals. For large enough samples like ours, where the input targets are identically selected, the average continuum from these masks should be a good match to the average continuum in the microshuffle masks, and we find this to be the case.  This continuum can then be subtracted from the averaged microshuffle slit, after a renormalization to the average counts\footnote{In practice we average the absolute value of the three pixels near the peak of each of the positive- and negative-flux spectra; the average of a nod-and-shuffle observation with perfect sky subtraction should be zero.} in the range $9000\mbox{\AA}<\lambda<9280\mbox{\AA}$.  The result is a two-dimensional image that contains only the residuals due to charge diffusion, as shown in the middle spectrum of Figure~\ref{fig-residuals}.
\begin{figure*}{
		\includegraphics[width=7.5in]{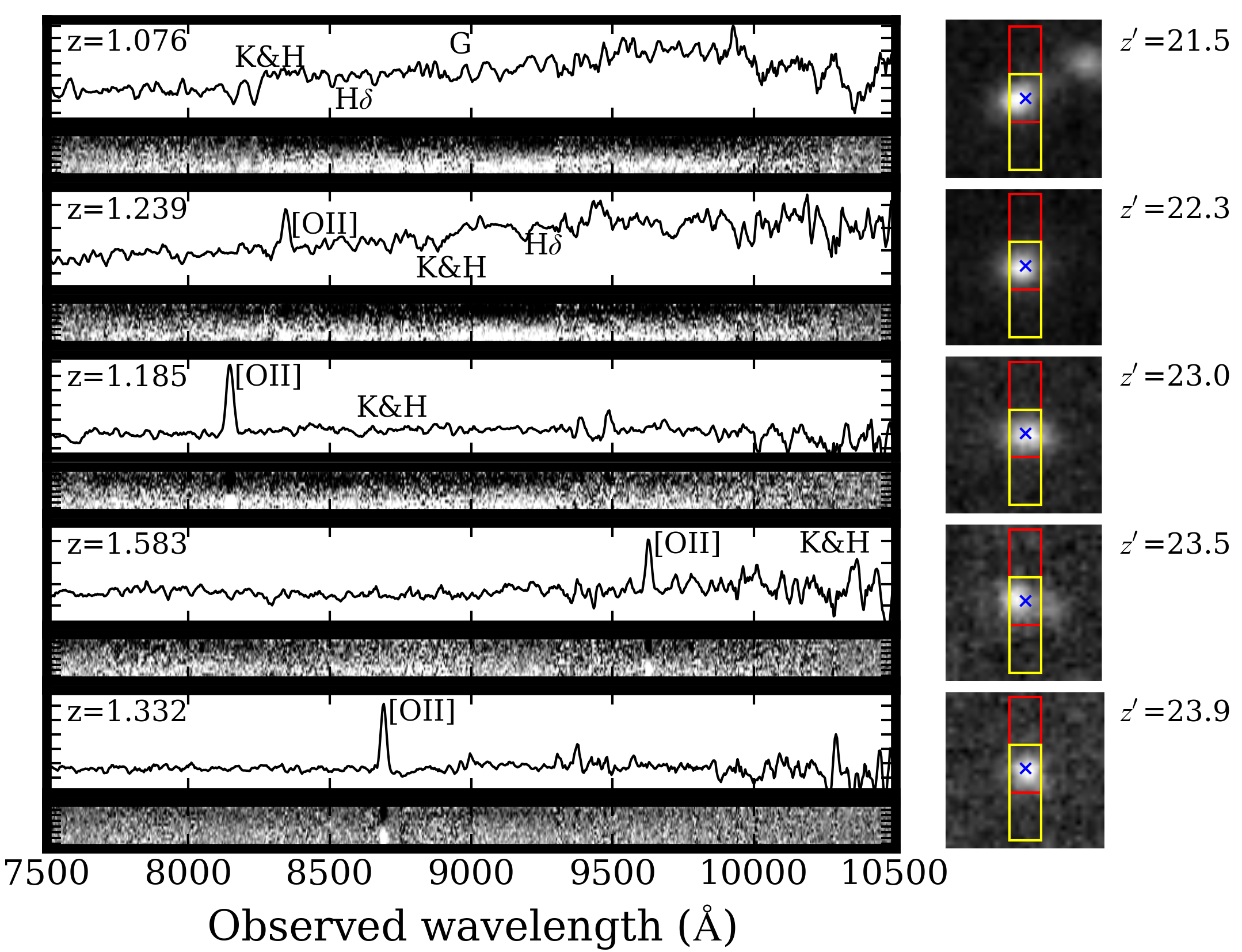}}
	\caption{Sample data are shown for five galaxies in a range of $z^{\prime}$ magnitudes, indicated by the numbers on the right.  Redshifts are given in the top left corner of each panel, and key identifiable spectral features are labelled (Calcium H and K, H$\delta$ and G-band absorption, and [OII] emission).  The stamps on the right show the $z^{\prime}$-band image, scaled to the minimum and maximum counts within the subimage.  The red and yellow rectangles show the predicted slit position in both nod positions.  To the left are the 1D (top) and 2D (bottom) reduced spectra, after all reduction including sky subtraction, charge diffusion correction, and telluric/sensitivity correction (the latter applied only to the 1D spectrum).  All spectra shown here are from 7.2ks of exposure, and the 1D spectra are convolved with a 5 pixel ($\sim 20$ \AA) boxcar filter.  For the faintest two galaxies, the goal is to build up the signal-to-noise by reobserving in multiple masks, with up to 54ks exposure by the end of the survey.\label{fig-samplespec}}
\end{figure*}

What remains is to subtract this ``master residual'' from the data, after appropriate rescaling.  The amplitude of the residual is expected to be directly proportional to the flux in the corresponding sky spectrum, since this light dominates over the object and uniformly fills the slit.  Thus we measure the mean signal of each column in the sky spectrum that corresponds to a given science slit.  Pixels with DQ flags, or sky values $<0$ or $>5000$, are masked.  The average is taken only of the four central rows, which are relatively free from science target flux.  We take the ratio of this average to the average of the same pixels in the combined sky spectrum, and use this to scale the master residual image.  Finally this scaled image is subtracted from the science data, at $\lambda>8000$\AA, where the effect is significant.  The resulting science spectrum is free from these residuals, as shown in the bottom panel of Figure~\ref{fig-residuals}.  The subtracted flux is stored in a new extension labelled 'REDFIX'.

Three masks taken in early 2015 (two in COSMOS-221 and one for COSMOS-125) had to be dealt with separately.  The charge diffusion here is larger than predicted by the simple scaling of the master residual described above\footnote{This may be related to a detector controller problem that was noted by Gemini a few weeks later, and which led to a much more severe charge smearing effect.}.    Having identified this, these images were excluded from the master residual described above.  To deal with these frames a similar process was followed, but using stacked residual frames in bins of date, $x_{ccd}$ and $y_{ccd}$.  With significant trial and error to choose appropriate bin sizes, corrections were found that work reasonably well for these three masks. 

The effect is also present on the EEV images from GMOS-N.  It does not present as much of a problem here, as the detector sensitivity has died off by the time the effect becomes most problematic, beyond $\sim 9600$\AA.  Since we do not plan to obtain any bandshuffle masks with the EEV detectors on GMOS-N, a different procedure is required to remove the continuum from the residual stack.  We take the bandshuffle continuum image described above and first rescale in the spatial direction to match the EEV pixel size, using a second order spline.  We then scale the intensity as a function of wavelength by the ratio of the EEV sensitivity relative to the Hamamatsu, using standard star observations.  This serves to adequately model the average EEV continuum image, at least for $\lambda>8000$\AA.  This is subtracted from the combined microshuffle data as for the Hamamatsu observations, to produce an appropriate master residual frame.

\subsection{Extraction and Flux Calibration}\label{sec-xtract}
We fully reduce each mask, including wavelength calibration, sky subtraction and (where necessary) charge diffusion correction, and then median combine the slits using {\it gemcombine}, first rejecting the lowest and highest pixels.  The reduced, two-dimensional image of each slit is 3\arcsec\ high, with an object spectrum at the top and an inverted spectrum at the bottom.  We first compute an average spatial profile of the slit, by computing the median within $8000\mbox{\AA}<\lambda<9750$\AA, rejecting bad pixels using the DQ mask.  We ignore the two pixels closest to either edge of the slit.  This profile is fit with two Gaussian distributions, one with amplitude $A$ and the other with amplitude $-A$.  Both are forced to have the same width, $\sigma$, and to be separated by a fixed amount given by the nod distance of $1\farcs45$. Next we repeat the process for small intervals of wavelength (typically 250 or 500 \AA), but keeping $\sigma$ fixed.  Thus we fit for only two parameters at each wavelength bin: the overall normalization, $A(\lambda)$, and the position of the bottom peak, $y(\lambda)$.  Finally, we plot $y(\lambda)$ as a function of $\lambda$, and fit a polynomial to it with 2$\sigma$ rejection.  The order of the polynomial, and the wavelength range of the fit, are determined interactively by the user.  Typically the order is 0--2, and the fit is done over $6500\mbox{\AA}<\lambda<9500$\AA.  

The spectral extraction is then a weighted sum of all pixels in a column (again omitting the top and bottom two pixels), where the weights are given by the double Gaussian function with vertical position at each wavelength given by the polynomial fit.  Thus most of the weight is given to the pixels at the centre of each spectrum.  The amplitude is irrelevant, but the sign of the function ensures that the spectrum with negative flux is subtracted from the one with positive flux.  

The extraction is made via a tool which shows the median extraction profile, the plot of $y(\lambda)$ vs $\lambda$ as well as the fit, the fit locations of both Gaussian peaks overlaid on the two dimensional spectrum, and the spectrum extracted from this fit.  Polynomial order, wavelength coverage and binning are chosen interactively to ensure a good fit to each spectrum.  The extraction parameters are stored in the header of the extracted spectrum.

Spectral flux standard observations are taken once each semester.  The standards are reduced using the same pipeline described above, including the QE correction for GMOS-S.  As these are not observed in nod-and-shuffle mode, however, sky subtraction is done classically by defining a sky region adjacent to the source.  The extracted spectrum is compared with tabluated values in the {\sc iraf} database to generate a sensitivity function which is then applied to the extracted science data.

Bands of telluric absorption at $6850<\lambda/\mbox{\AA}\sim6940$ (B band), $7550<\lambda/\mbox{\AA}<7710$ (A-band), $8120<\lambda/\mbox{\AA}<8370$ and $8940<\lambda/\mbox{\AA}<9840$ are corrected using an {\sc iraf} package that cross correlates telluric features from our standard stars to compute a shift and scale factor before subtracting from the data. 
This does not provide a perfect correction for the red features ($\lambda>8000$\AA), as these lines vary on short timescales.  Starting in 2016B we have been including one bright star in each mask.  By fitting templates to those stars we expect to derive a telluric correction which is applicable to all spectra in that mask.  For earlier observations, we are exploring ways to use the existing data to derive improved corrections.
\begin{figure*}{
		
		\includegraphics[clip=true,trim=0mm 0mm 0mm 0mm,width=7in,angle=0]{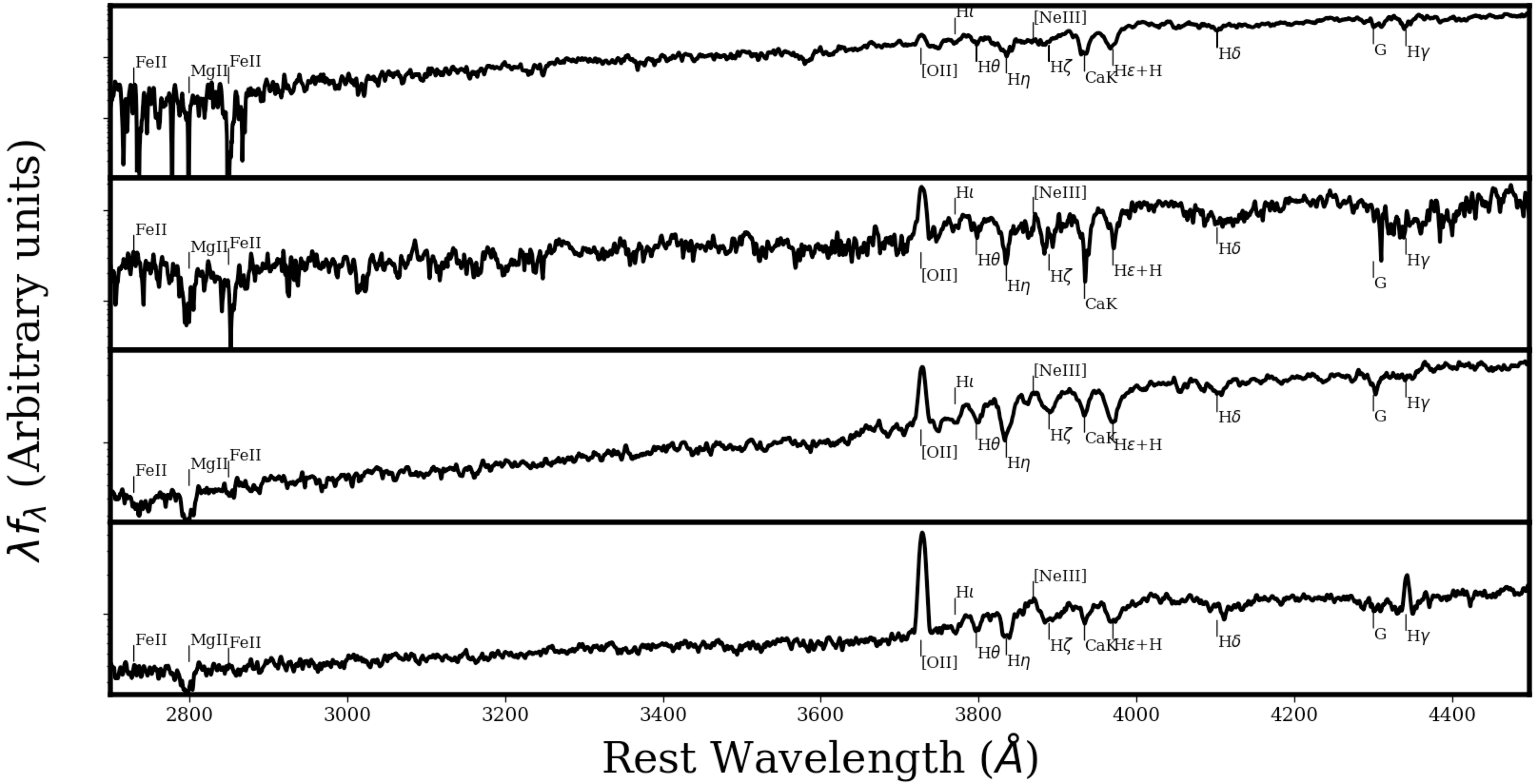}
	}
	\caption{Averaged, resampled spectrum obtained by combining all of our $1.0<z<1.5$ spectra with good redshifts.  From top to bottom the panels show a) bright, red; b) faint, red; c) bright, blue; d) faint, blue.  The colour division is made based on D4000, and the bright/faint division is the subsample median.\label{fig-stack}}
\end{figure*}

\subsection{Final spectra and science analysis}
In Figure~\ref{fig-samplespec} we show some sample images and extracted 1D and 2D spectra for five targets after just 7.2ks exposure.  A range of magnitudes are shown; the faintest galaxies ($z^{\prime}>23.5$) will ultimately accrue up to 54ks of exposure by observing in multiple masks.  The 2D spectra shown are sky subtracted and corrected for charge diffusion.  In these images, there is a positive and negative-flux copy of the spectrum due to the nod-and-shuffle technique.  Note the spectrum is free from sky line residuals.  The 1D spectra are extracted as described in \S~\ref{sec-xtract}, including sensitivity calibration and preliminary telluric correction.  Four of the spectra shown here show a strong [OII] emission line, clearly identifiable in both the 1D and 2D spectrum.  The top spectrum is a pure absorption line system, with the Ca H+K lines easily identifiable at $\lambda\sim8200$\AA.

Preliminary redshifts are being determined independently using the {\it runz} code\footnote{http://www.physics.usyd.edu.au/~scroom/runz/zguide.html}, and an updated version of the DEEP2 {\it spec1d} pipeline \citep{DEEP2_data}.  Future improvements will include adding rest-UV templates generated from our own data.  

Stellar masses for the sample will be derived from SED fitting to multiwavelength photometry, including deep [3.6]$\mu$m imaging.  For those clusters which currently lack sufficiently deep data,  we have shown \citep{GCLASS12} that corrections for $M/L$ based on D4000 are sufficient to obtain masses to within a factor $\sim 2$ of those derived from SED--fitting.  This requires D4000 to be measured to within 20\% precision (corresponding to S/N$\sim$0.7 per pixel), which will be achievable for every galaxy for which we can get a redshift. 

To measure the quiescent fraction we need to classify our galaxies.  This is best done using colour--colour diagrams spanning the rest frame NUV to NIR, which does an excellent job of separating dusty star--forming galaxies from truly passive galaxies \citep[e.g.][]{Muzzin_SMF,Mok13,Arnouts+13}.  This requires deep imaging spanning $u$ through $K$, which we have now obtained for most clusters in our sample (see \S~\ref{sec-mwi}).

Good age estimates for young to intermediate age populations can be obtained from absorption lines of H$\delta$, Calcium K and G-band \citep{Muzzin14}.  While these lines cannot generally be measured reliably from individual spectra, we only need to stack $\sim 50$ galaxies to obtain $S/N>20$ per resolution element, sufficient to derive meaningful ages \citep[e.g.][]{Conroy14}.  Our sample size therefore allows us to calculate average spectra in bins of stellar mass, radius and halo mass, and constrain the luminosity--weighted age to within $\sim 20$ per cent \citep{Demarco+10,Muzzin14,Mok14}. As an early example of this, Figure~\ref{fig-stack} shows a combined spectrum of all our $1.0<z<1.5$ spectra with good redshifts as of the end of the 2016A semester.  We divide the sample into red and blue based on the D4000 index, with $D4000>1.5$ being red.  Within each subsample we divide into two equal-sized luminosity bins. Each stack is comprised of about 40 galaxies.  As well as the usual strong absorption (Ca H+K, G-band) and emission ([OII]) lines, numerous weaker features are also detected at high S/N, including FeII, MgII, [NeIII] and higher order Balmer lines.  By modelling the spectra with stellar population synthesis models, we will be able to determine ages and metallicities for galaxies as a function of their stellar mass and environment.

For the dynamical analysis, we expect to have $>50$ confirmed members in all systems but the groups, sufficient to keep within $\pm 10$\% the average bias of total mass estimates from their velocity dispersions  \citep{Biviano+06}.  
More accurate, precise and detailed dynamics can be determined when galaxies from several clusters are combined.    For example, the 500 cluster members expected for our SpARCS sample alone are sufficient to constrain both the average total
mass radial profile $M(r)$, and also the
velocity anisotropy profile $\beta(r)$ of their
member galaxies.  Importantly, the sample size will be large enough to do this separately for the passive and star-forming
populations, which are known to have different kinematics \citep[e.g.][]{Mohr+96,Biviano+97}.  We will achieve this using the MAMPOSSt technique \citep{MBB13,Biviano+13}
which breaks the intrinsic degeneracy
between $M(r)$ and $\beta(r)$ in the Jeans equation. We will combine this analysis with 
the complementary caustic technique \citep{DG97}, to construct solutions that are independent of assumptions about dynamical equilibrium.   
\begin{figure}{
		\includegraphics[clip=true,trim=0mm 0mm 0mm 0mm,width=3.5in,angle=0]{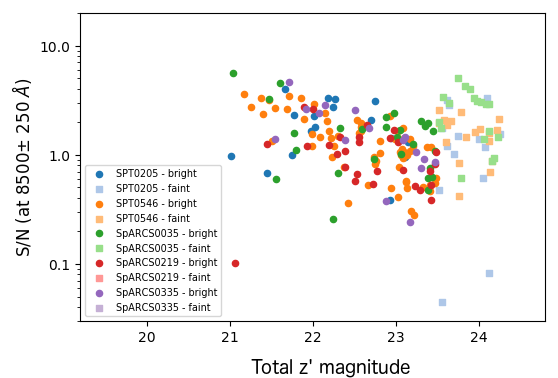}}
	\caption{The projected final S/N per \AA, measured in a 500\AA\ window centred on 8500 \AA, is shown for every primary target spectrum obtained as of semester 2016B, as a function of total $z^{\prime}$-band magnitude.  
		Each symbol colour corresponds to a different mask.  The measured S/N for each spectrum is scaled to its predicted value at the end of the survey, assuming $3$h exposures for $z^{\prime}<23.5$ galaxies, and $15$h exposures for $z^{\prime}>23.5$ galaxies.  \label{fig-snr2}}
\end{figure}
\begin{figure*}{
		\includegraphics[clip=true,trim=0mm 0mm 0mm 0mm,width=3.2in,angle=0]{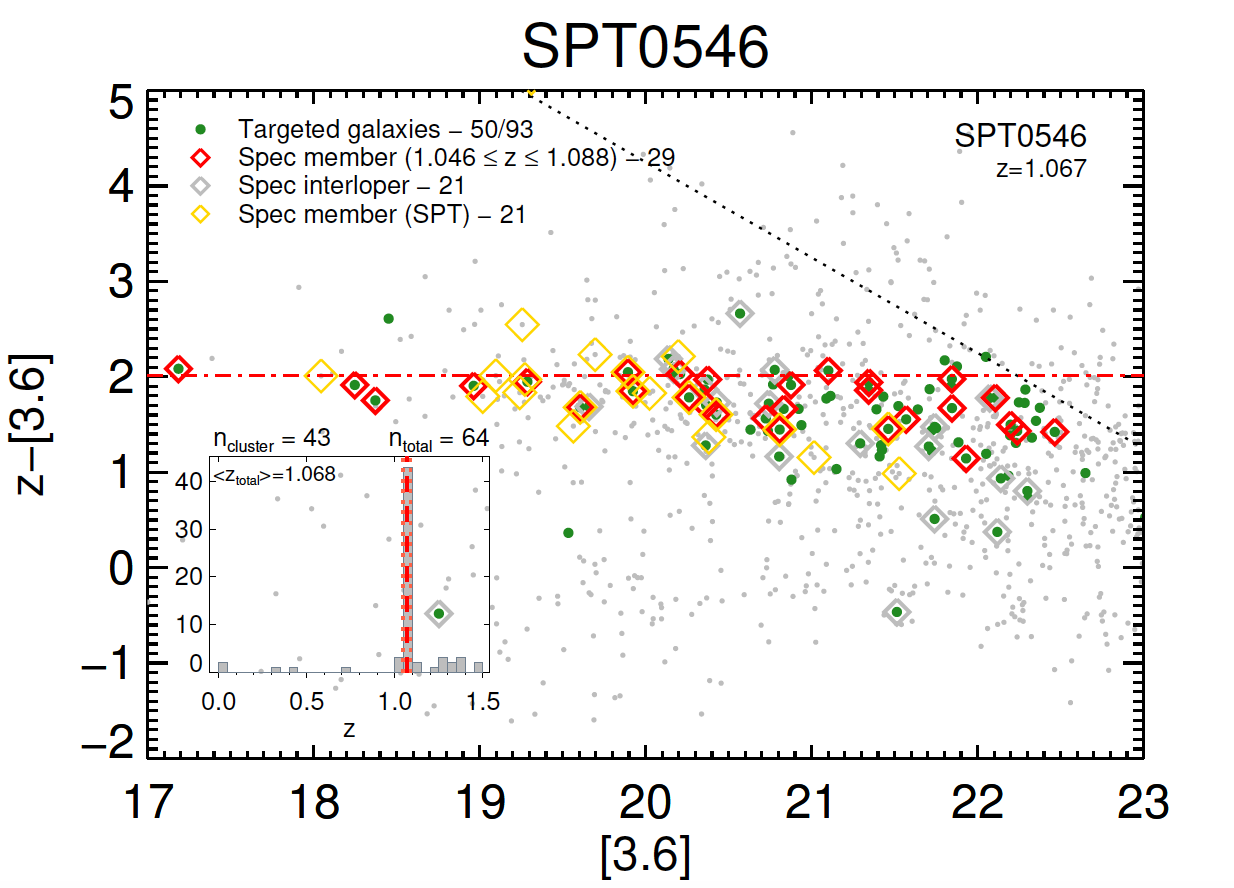}\includegraphics[clip=true,trim=0mm 0mm 0mm 0mm,width=2.35in,angle=0]{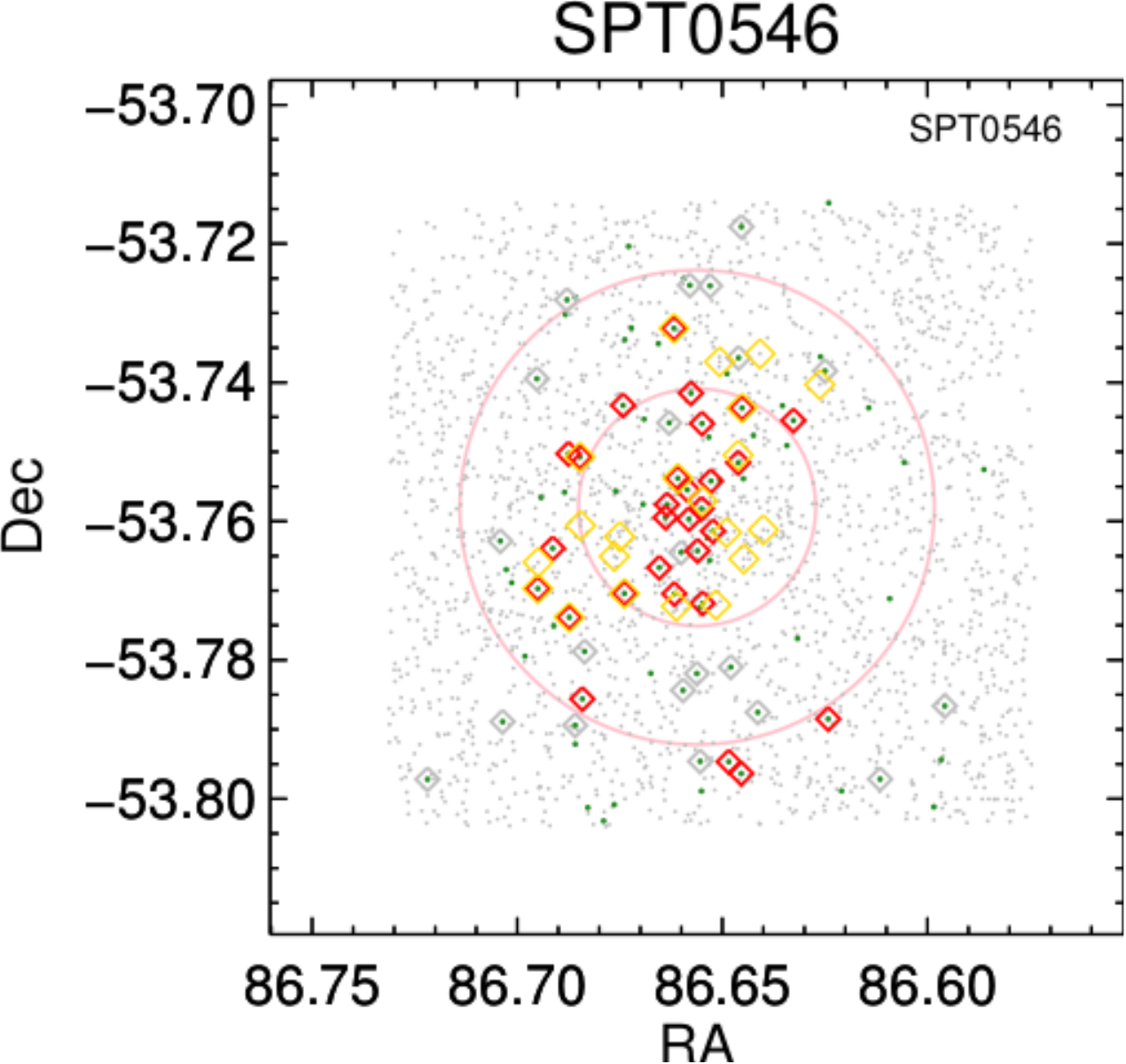}
		\includegraphics[clip=true,trim=0mm 0mm 0mm 0mm,width=3.2in,angle=0]{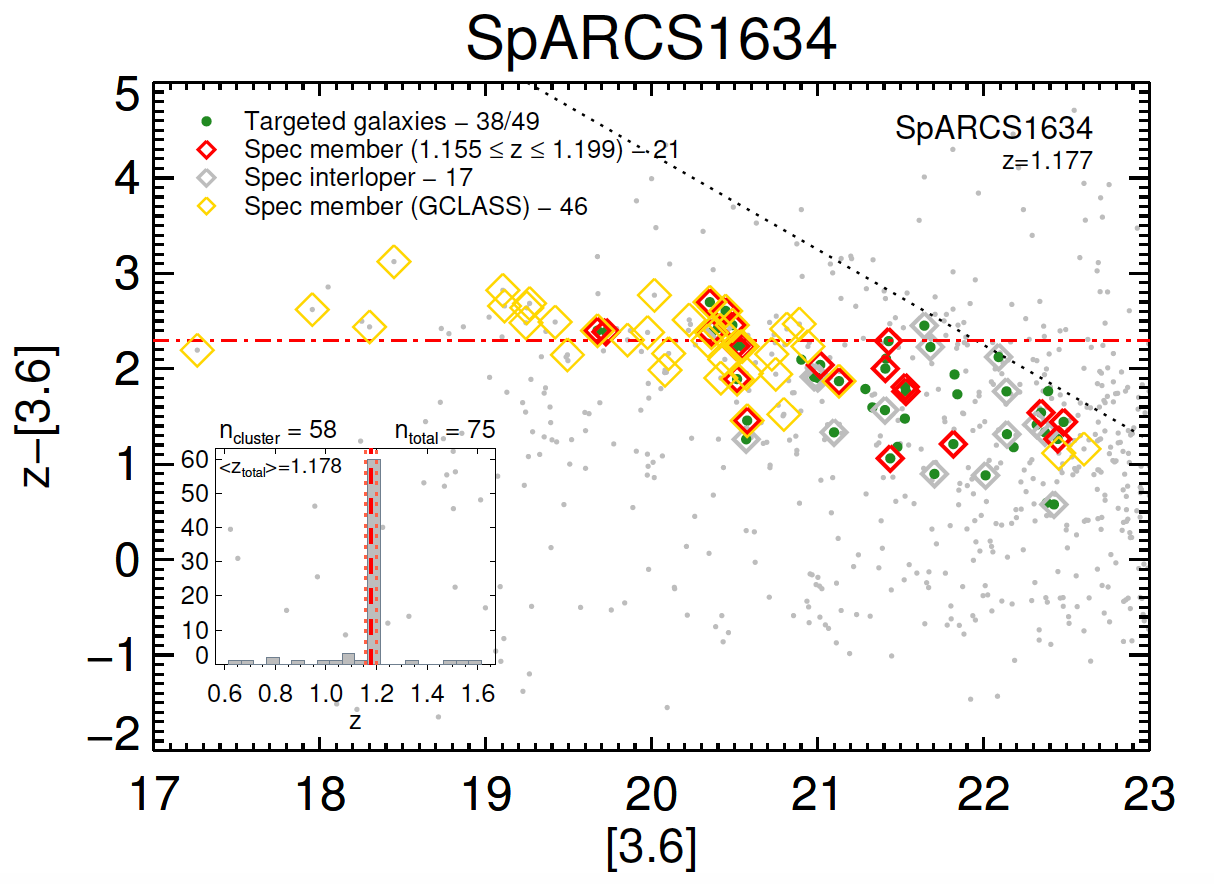}\includegraphics[clip=true,trim=0mm 0mm 0mm 0mm,width=2.35in,angle=0]{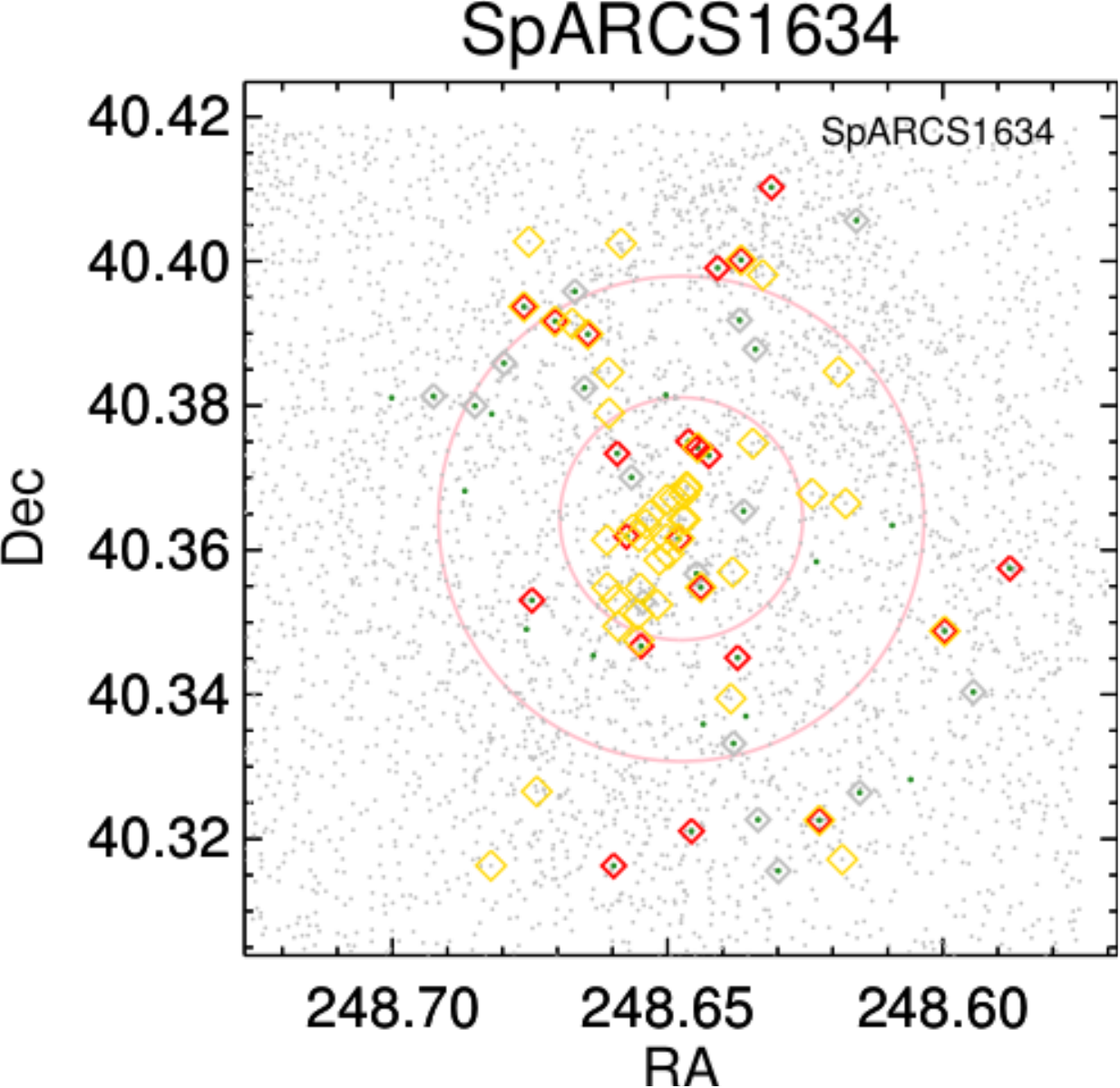}
	}
\caption{Preliminary colour magnitude diagrams (left) and spatial distribution of the spectroscopy (right) are shown for two of our clusters, as labelled.  The red horizontal line shows the adopted red sequence position, and the dotted black line represents the $z^{\prime}$ target selection limit.  Red diamonds indicate new spectroscopic members from our GOGREEN spectroscopy, while yellow diamonds are existing redshifts from GCLASS and SPT.  Grey diamonds are GOGREEN galaxies with redshifts confirmed to be non-members, while green circles are galaxies that do not yet have a reliable redshift.  Neither field has achieved its full S/N at this point in the survey, so redshift completeness will increase.  The inset on the left plot shows the redshift histogram with an estimated number of cluster members and mean redshift indicated.  The spatial plots on the right show how the spectroscopic members are distributed relative to 0.5Mpc and 1.0Mpc radii around the field centre.\label{fig-cmd}}
\end{figure*}

\section{Survey Status}\label{sec-status}

The total time awarded to GOGREEN was 438.3h, and a specific allocation is made each semester. 
Through mid-2017A, we have successfully executed 242h of the 338.9h allocated in this way, for a 71\% completion rate.
100\% of the required deep $z^{\prime}$-band imaging was obtained in the first two semesters, as planned; our spectroscopic program is now just over half complete.
Our program was hit particularly hard by the very bad weather at Gemini-S during 2015, when only 18.3 of our 69h allocation could be executed.  This low completion rate has required an adjustment to our overall strategy.  In early 2016 it was decided to postpone any further observations of the nine group targets in COSMOS and SXDS and focus on completing the massive cluster sample, to ensure an impactful program.

The other obstacle was the delayed deployment of Hamamatsu detectors on Gemini-N, which meant that observations of our high redshift northern targets were pushed toward the end of the program.  For these reasons we requested, and were granted, a program extension of two semesters, to the end of 2018A.

Analysis of the spectroscopic data acquired so far shows we are reaching our target S/N.  
In Figure~\ref{fig-snr2} we show the projected, final S/N for all objects for which we have some existing data.  The S/N is measured over 8250--8750\AA, the most relevant range for redshift determination.  The measurements are scaled from the current exposure time for each spectrum, assuming a final exposure of 3h for $z^{\prime}<23.5$ galaxies and 15h for $z^{\prime}>23.5$ galaxies.  This shows that we expect to achieve S/N$\gtrsim$0.7 per \AA, or S/N$>3$ per resolution element ($\sim 20$\AA) for  $\sim 80$ per cent of our targets.  This is consistent with our proposal objectives.

In Figure~\ref{fig-cmd} we show two examples of colour-magnitude diagrams, with coloured points indicating the spectroscopic sample.  GOGREEN is extending existing spectroscopy in these systems as expected, down to [3.6]$\mu$m$<22.5$, with high completeness.  Neither of these fields has achieved its full S/N at this point; thus, redshift completeness will increase by the end of the survey. 

In the right panels of Figure~\ref{fig-cmd} we show the spatial distribution of the galaxies in these two clusters.  Unlike GCLASS, GOGREEN does not focus on the dense core of the cluster, but aims for a more even sampling beyond a $\sim 1$ Mpc radius.  However, the restriction to one GMOS field of view, and the need to keep the full wavelength coverage in most spectra, means the spectral sample is extended in one dimension more than the other.  In clusters which do not have existing spectroscopy of the core as these examples do, we execute at least one band-shuffled mask to ensure a high sampling of relatively bright galaxies near the centre.

\subsection{Public Data Release}
GOGREEN has committed to release the first data products no later than one year after the end of the survey.  The final data release will include at least reduced spectra, reduced GMOS images, and catalogues of redshifts, GMOS photometry and advanced data products including line indices and photometric redshifts.  The GCLASS and GEEC2 data will also be provided as part of this release.  Details will be available in a forthcoming paper devoted to the data release.

\section{Conclusion}
The GOGREEN survey is a Large Program on Gemini North and South, using a large allocation of time ($>400$h) to construct an unprecedented sample of homogeneously selected galaxy spectroscopy in 21 galaxy clusters and groups at $1.0<z<1.5$.  The targets are chosen to span a wide range in halo mass, such that they correspond to the progenitors of the massive clusters and groups that are well studied at $z\sim 0$.  The red sensitivity of the Hamamatsu detectors, coupled with the nod-and-shuffle mode, allows good quality spectra to be obtained at $\lambda < 1.05 \mu$m, and with this we probe galaxies of all types with $M_\ast\gtrsim10^{10.3}M_\odot$ over the whole redshift range.  
We provide a thorough description of the GMOS data reduction, and the specific challenges associated with the Hamamatsu detectors.    In addition to the deep spectroscopy, we have acquired deep multiwavelength photometry, spanning $ugrizYJK$ and Spitzer [3.6]$\mu$m in most systems, over and beyond the full spectroscopic field of view.  With these data we will investigate the role of environment in galaxy evolution, at an epoch when the overall galaxy population is forming stars at a much higher rate than today.  The data also enable an analysis of cluster dynamics and stellar content over a wide range in halo mass.  We anticipate the sample will have high legacy value, including a sample of $\sim 600$ field galaxies with spectroscopic redshifts down to faint magnitudes $z^{\prime}<24.25$.  The first public data release will occur within one year of the survey completion.

\section{Acknowledgments}
We extend our thanks and appreciation to Gemini, and the Gemini communities, for supporting this Large Program.  We also thank Gemini for their support of junior observers on several of our observing runs.  
\par
This research is supported by the following grants:  NSERC Discovery grants (MLB and LCP); Universidad Andr{\' e}s Bello internal project grants DI-651-15/R and DI-18-17/RG (JN);  NSF grants AST-1517815 \&  AST-1211358 (GHR), AST-1517863 (GW) and AST-1518257 (MCC); NASA, through grants AR-14310.001 \& GO-12945.001-A (GHR), GO-13306, GO-13677, GO-13747 \& GO-13845/14327 (GW),  AR-13242 \& AR-14289 (MCC) and HST-GO-14734 (AW) from the Space Telescope Science Institute, which is operated by AURA, Inc., under NASA contract NAS 5-26555; STFC (SLM); the BASAL Center for Astrophysics and Associated Technologies (CATA), and
FONDECYT grant N. 1130528 (RD); the European Research Council under FP7 grant number 340519 (RFJvdB); the National Research Foundation of South Africa (DGG);  Chandra grant AR6-17014B (AFF);  and FONDECYT postdoctoral research grant no 3160375 (PC).   AW was also supported by a Caltech-Carnegie Fellowship, in part through the Moore Center for Theoretical Cosmology and Physics at Caltech.  GHR also acknowledges the hospitality and financial support of the Max-Planck-Institute for Extraterrestrial Physics, the European Southern Observatory and the International Space Sciences Institute as well as the hospitality of the Max-Planck-Institute for Astronomy and Hamburg Observatory. 

This paper includes data gathered with the Gemini Observatory, which is operated by the Association of Universities for Research in Astronomy, Inc., under a cooperative agreement with the NSF on behalf of the Gemini partnership: the National Science Foundation (United States), the National Research Council (Canada), CONICYT (Chile), Ministerio de Ciencia, Tecnología e Innovación Productiva (Argentina), and Ministério da Ciência, Tecnologia e Inovação (Brazil); the 6.5 meter Magellan Telescopes located at Las Campanas Observatory, Chile;  the Canada-France-Hawaii Telescope (CFHT) which is operated by the National Research Council of Canada, the Institut National des Sciences de l'Univers of the Centre National de la Recherche Scientifique of France, and the University of Hawaii; MegaPrime/MegaCam, a joint project of CFHT and CEA/DAPNIA; Subaru Telescope, which is operated by the National Astronomical Observatory of Japan; and the ESO Telescopes at the La Silla Paranal Observatory under programme ID 097.A-0734.

\par
\noindent
$^{1}$Department of Physics and Astronomy, University of Waterloo, Waterloo, Ontario, N2L 3G1, Canada\\
$^{2}$South African Astronomical Observatory, PO Box 9, Observatory, Cape Town, 7935, South Africa\\
$^{3}$Centre for Space Research, North-West University, Potchefstroom 2520, South Africa\\
$^{4}$Department of Physics and Astronomy, York University, 4700 Keele Street, Toronto, Ontario, ON MJ3 1P3, Canada\\
$^{5}$Department of Physics and Astronomy, The University of Kansas, Malott room 1082, 1251 Wescoe Hall Drive, Lawrence, KS 66045, USA\\
$^{6}$Department of Physics and Astronomy, University of California, Irvine, 4129 Frederick Reines Hall, Irvine, CA 92697, USA\\
$^{7}$Australian Astronomical Observatory, 105 Delhi Road, North Ryde, NSW 2113, Australia\\
$^{8}$INAF-Osservatorio Astronomico di Trieste, via G. B. Tiepolo 11, 34143, Trieste, Italy\\
$^{9}$Departamento de Astronom\'ia, Universidad de Concepci\'on, Casilla 160-C, Concepci\'on, Regi\'on del Biob\'io, 0000-0003-3921-2177, Chile\\
$^{10}$School of Physics and Astronomy, University of Birmingham, Edgbaston, Birmingham B15 2TT, England\\
$^{11}$Departamento de Ciencias F\'isicas, Universidad Andres Bello, Fernandez Concha 700, Las Condes 7591538, Santiago, Regi\'on Metropolitana, Chile\\
$^{12}$MIT Kavli Institute for Astrophysics and Space Research, 70 Vassar St, Cambridge, MA 02109, USA\\
$^{13}$Department of Astronomy and Astrophysics, University of Toronto 50 St. George Street, Toronto, Ontario, M5S 3H4, Canada\\
$^{14}$Department of Physics and Astronomy, University of California Riverside, 900 University Avenue, Riverside, CA 92521, USA\\ 
$^{15}$European Southern Observatory, Alonso de Cordova 3107, Vitacura, Casilla 19001, Santiago de Chile, Chile.\\
$^{16}$Laboratoire AIM, IRFU/Service d'Astrophysique - CEA/DRF - CNRS - Universit\'e Paris Diderot, B\^at. 709, CEA-Saclay, 91191 Gif-sur-Yvette Cedex, France\\
$^{17}$Steward Observatory and Department of Astronomy, University of Arizona, Tucson, AZ, 85719\\
$^{18}$Observatorio Astron\'omico de C\'ordoba (UNC) and Instituto de Astronom\'ia Te\'orica y Experimetal (CONICET-UNC). C\'ordoba, Laprida 925, Argentina.\\
$^{19}$Institute for Computational Cosmology, Durham University, South Road, Durham DH1 3LE UK\\
$^{20}$Center for Extra-galactic Astronomy, Durham University, South Road, Durham DH1 3LE UK\\
$^{21}$Department of Physics, University of Helsinki, Gustaf H\"allstr\"omin katu 2a, FI-00014 Helsinki, Finland\\ 
$^{22}$Max-Planck-Institut f\"ur extraterrestrische Physik, 
Giessenbachstr. 1, Garching, D-85741, Germany\\
$^{23}$Department of Physics and Astronomy, McMaster University, Hamilton ON L8S 4M1, Canada\\
$^{24}$IPAC, Caltech, KS 314-6, 1200 E. California Blvd, Pasadena, CA 91125, USA
$^{25}$Carnegie Observatories, Pasadena, CA, USA\\
$^{26}$TAPIR, California Institute of Technology, Pasadena, CA, 91125, USA
$^{27}$Department of Physics, University of California, Davis, CA, 95616, USA

\bibliography{ms}
\end{document}